\documentstyle[12pt]{article}
\begin{document}

\title{The SUSY EW-like corrections to top pair production in photon-photon
       collisions}

\vspace{-3mm}
\author{{ Zhou Mian-Lai$^{b}$, Ma Wen-Gan$^{a,b,c}$, Han Liang$^{b}$,
          Jiang Yi$^{b}$ and Zhou Hong$^{b}$}\\
{\small $^{a}$CCAST (World Laboratory), P.O.Box 8730, Beijing 100080, China} \\
{\small $^{b}$Modern Physics Department, University of Science and}\\
{\small Technology of China, Anhui 230027, China}  \\
{\small $^{c}$Institute of Theoretical Physics, Academia Sinica,
        P.O.Box 2735, Beijing 100080, China.}}

\date{}
\maketitle
\vspace{-8mm}

\begin{abstract}
We studied the one-loop contributions of the gaugino-Higgsino-sector
to the process of top-pair production via $\gamma \gamma$ fusion at NLC
in frame of the Minimal Supersymmetric Model(MSSM). We find that the
corrections to $\gamma \gamma \rightarrow t\bar{t}$ and
$e^+ e^- \rightarrow \gamma \gamma \rightarrow t\bar{t}$ are found
to be significant and can approach to a few percent and one percent,
respectively. Furthermore, the dependences of the corrections on the
supersymmetric parameters are also investigated. The corrections are not
sensitive to $M_{SU(2)}$ (or $|\mu|$) when $M_{SU(2)}~>>~|\mu|$
(or $|\mu|~>>~M_{SU(2)}$) and are weakly dependent on the $\tan{\beta}$
with $M_Q$ (or $|\mu|$) being large enough. But they are sensitive to the
c.m.s. energy of the incoming photons.\\
\end{abstract}
{\large\bf PACS: 12.15.Lk, 12.60.Jv, 12.60.Cn, 12.60.Fr, 14.65.Ha}

\vfill \eject


\noindent{\large\bf I. Introduction}
\par
   The direct discovery of the top quark was presented in 1995 by the CDF
and D0 experiments at the Fermilab Tevatron\cite{a1}. This is considered
to be a remarkable success for the Standard Model(SM),
since the present value of the top mass determined as PDG average is
$173.8\pm 5.2~GeV$ from the direct observation of top events\cite{aa1},
which coincides with the indirect
determination from the available precise data of electroweak experiments.
But the SM has still some theoretical problems, like the hierarchy
problem, the necessity of fine tuning and the non-occurrence of gauge
coupling unification at high energies. The Supersymmetric Models(SUSY)
can solve these problems by presenting an additional symmetry. Among all the
extensions of the SM, the Minimal Supersymmetric Standard Model (MSSM)\cite{a6}
is the most attractive one at present, since it is the simplest case of the
SUSY models.
\par
Due to the strong Yukawa couplings of top quark, the SUSY
electroweak radiative corrections in top-pair production process are
specially interesting. People believe that the accurate measurement of top
quark pair production at the present and future colliders, should be effective
in measuring the physical effects induced by the virtual supersymmetric
particles and can afford us much information about the MSSM. Any deviation of
the cross section of top-pair production from the SM predictions, including
QCD and electroweak radiative corrections, would give a hint of new physics
beyond the SM. Therefore testing this process to make the indirect search for
virtual SUSY particles, is an attractive theme at present and future colliders.
\par
In previous studies, many works were concentrated on the top-pair
production at the $e^+e^-$ and hadron colliders, such as the LEP2, LHC and
Tevatron. In references\cite{a6a}, the SUSY QCD and SUSY
electroweak-like (EW-like) corrections at $pp$ colliders are presented.
Recently, W. Hollik and C. Schappacher calculated the MSSM radiative one-loop
corrections to top-pair production via $e^+e^-$ collisions at LEP2 energies
and found the relative difference between the predictions of the MSSM and the
SM is typically below $10\%$\cite{a6b}.
\par
The future Next Linear Collider(NLC) is designed to give the facilities for
both $e^+ e^-$ and $\gamma \gamma$ collisions at the energy of $500
\sim 2000~GeV$ with a luminosity of the order $10^{33} cm^{-2} s^{-1}$
\cite{a2}. A large number of top quark and other particle pairs can be
produced at this machine operating in $\gamma \gamma$ collision mode with an
agreeable production rate \cite{a3}. The events would be much cleaner
than those produced at $pp$ and $p \bar{p}$ colliders. It has been also
found that the $t\bar{t}$ production rate in $\gamma\gamma$ collisions
is much larger than that from the direct $e^{+}e^{-} \rightarrow t\bar{t}$
production both with and without considering the threshold QCD effect of
top quark pair at center-of-mass energies of the electron-positron
system around $1~TeV$\cite{a4}. Thus the process $\gamma\gamma \rightarrow t
\bar{t}$ has a large potential for studying top physics directly.
\par
The next-to-leading order QCD corrections in the SM and MSSM for this process
both for polarized and unpolarized photon-photon collisions have been discussed
in detail in Ref.\cite{a5}. There it was shown that the QCD corrections in
both the SM QCD and the MSSM QCD are about $10 \%$ and of the order
$-10^{-2}$, respectively. A. Denner, S. Dittmaier and M. Strobel calculated
the corrections to the process $\gamma \gamma \rightarrow t \bar{t}$ in the
electroweak standard model and found that the correction reduction
for unpolarized or equally polarized photons can reach almost
$10\%$ close to threshold \cite{a5q}. In the reference\cite{a5p},
C.S. Li et al. calculated the $O(\alpha m_t^2/m_W^2)$ Yukawa corrections
from Higgs-sector to top pair production via photon-photon collision in
the SM, the general two-Higgs-doublet model (2HDM) as well as the MSSM.
They found that the correction to the cross section is about a few
percent in the SM, but the correction can be more significant ($>$10\%)
in the MSSM. Therefore the SUSY loop contributions have considerable effects.
In this paper, we study the possible effects from the additional
EW-like one-loop corrections through the virtual presence of
charginos, neutralinos and squarks at the NLC. We provide explicit analytical
expressions for the form factors which parametrize the one-loop corrections
of $\gamma \gamma \rightarrow t \bar{t}$ subprocess, and present numerical
results both for the subprocess and process $e^+ e^- \rightarrow \gamma\gamma
\rightarrow t \bar{t}$ at the NLC.
\par
The paper is organized as follows: In
Sec. II, the theory about the chargino/neutralino is introduced, and
the relative Feynman rules used in the calculation are listed. In Sec. III,
we discuss the tree-level and one-loop EW-like correction cross section,
respectively, and give the explicit analytical formulae for them.
In Sec. IV, the numerical results and discussions are described. Finally,
we give a short summary. In the appendix, the form factors used in the
cross section calculations are listed in detail.\\

\begin{flushleft} {\bf II. Lagrangian and Feynman Rules}
\end{flushleft}
\par
We denote the process of the top-pair production via $\gamma\gamma$ fusion as
$$
\gamma (p_3) \gamma (p_4) \longrightarrow \bar{t}(p_1) t(p_2),
\eqno{(2.1)}
$$
where $p_{1,2}$ and $p_{3,4}$ represent the four-momenta
of the outgoing top quark pair and the incoming photons, respectively.
In this work, we consider one-loop corrections of the gaugino-Higgsino-sector
in the MSSM to this process. At the one-loop EW-like
correction order, the vertex $\gamma t\bar{t}$ is modified by the virtual
exchange of two charginos $\tilde{\chi}^{\pm}_{i=1,2}$ and four neutralinos
$\tilde{\chi}^{0}_{i=1 \sim 4}$, which are respectively combinations of charged
gaugino and Higgsino (for charginos), and neutral gaugino, Higgsino, photino
and zino (for neutralinos).  The mass eigenstates $\tilde{\chi}^{0}_{1,2,3,4}$,
$\tilde{\chi}^{\pm}_{1,2}$ for the charginos and neutralinos are respectively
obtained by diagonalizing the mass matrices X and Y in four component
representation.\cite{a6}\cite{a7}. The chargino mass term in lagrangian has
the form:
$$
 {\cal L}_{m}=-\frac{1}{2}(\psi^{+}~ \psi^{-}) \left(
      \begin{array}{ll} 0  &  X^{T}  \\ X  &  0
      \end{array}
      \right) \left( \begin{array}{l} \psi^{+}  \\
      \psi^{-} \end{array} \right),
\eqno{(2.2)}
$$
with $2\times2$ X defined in reference\cite{a6}.
The two masses of chargino $m_{\tilde{\chi}^{+}_{1,2}}$ extracted from
the diagonal elements of matrix X are worked out as
\begin{eqnarray*}
      M_{\pm}^{2} &=& \frac{1}{2}
        \left\{ M_{SU(2)}^{2} + \mu^{2} + 2 m_{W}^{2} \pm
        \left[ (M_{SU(2)}^{2}-\mu^{2})^{2} + 4 m_{W}^{4} \cos^{2}2\beta +
        \right. \right. \\
&& ~~~~~\left. \left. 4 m_{W}^{2} (M_{SU(2)}^{2} +
         \mu^{2}+ 2 M_{SU(2)} \mu \sin 2\beta )
                     \right]^{1/2} \right\}, ~~~~~~~~~~~~~~~~~~~~~~~~~~(2.3)
\end{eqnarray*}
\par
As to the neutralino sector, the mass term in lagrangian has the form as:
$$
 {\cal L}_{m}=-\frac{1}{2}(\psi^{0})^{T} Y \psi^{0} + h.c.,
\eqno{(2.4)}
$$
The definition of the $6 \times 4$ matrix Y can also be found in \cite{a6}
\cite{a7}.
\par
Since we do not take the CP violation into account, so all the possible
CP phases\cite{a7} are assumed to be zero. The physical masses of neutralinos
are obtained by utilizing the transformation matrix $N$ to diagonalize the
$4 \times 4$ mass matrix Y. The detailed steps to work out $N$ and the
diagonal matrix $Y_{D}$ are described in reference \cite{a7}. The above
equations show
that the chargino and neutralino masses are related to the MSSM parameters
$M_{SU(2)}$, $M_{U(1)}$, $\mu$ and $\tan{\beta}$. In our work we adopt the
assumption that the $SU(2) \times U(1)$ theory is embedded in grand unified
theory (GUT), so we have the following relation:
$$
  M_{U(1)}=\frac{5 s_W^2}{3 c_W^2} M_{SU(2)}.
\eqno{(2.5)}
$$
\par
In the MSSM, each quark has two scalar partners called squarks:
$\tilde{q}_L$ and $\tilde{q}_R$. Without considering CP phases,
the mass matrix of scalar quark takes the following form:\cite{s9}
$$
 -{\cal L}_{m}=\left( \begin{array}{ll}
        \tilde{q}^{\ast}_{L} &  \tilde{q}^{\ast}_{R}
              \end{array} \right)
       \left( \begin{array}{ll}  m^2_{\tilde{q}_{L}} &  a_{q} m_{q} \\
           a_{q} m_{q}  &  m^2_{\tilde{q}_{R}}
              \end{array}  \right)
       \left( \begin{array}{ll}  \tilde{q}_{L}  \\  \tilde{q}_{R}
              \end{array}  \right),
\eqno{(2.6)}
$$
The expressions of the masses for the squark current eigenstates are listed
in Appendix A Eqs.$(A.1)\sim(A.3)$.
Then the masses of $\tilde{q}_1$ and $\tilde{q}_2$ read
$$
(m^2_{\tilde{q}_1},m^2_{\tilde{q}_2})=\frac{1}{2} \{
     m^2_{\tilde{q}_L} + m^2_{\tilde{q}_R} \mp [ (m^2_{\tilde{q}_L} -
     m^2_{\tilde{q}_R})^2 + 4 a_q^2 m_q^2 ] ^{\frac{1}{2}} \}.
\eqno{(2.7)}
$$
\par
The Feynman rules for the couplings of $t-\tilde{b}_{L,R}-
\tilde{\chi}^{+}_{1,2}$ and $t-\tilde{t}_{L,R}-\tilde{\chi}^{0}_{1,2,3,4}$
are presented in Ref.\cite{a6}. The squark mixing angles $\theta_{\tilde{b}}$,
$\theta_{\tilde{b}}$ and phases $\phi_{\tilde{b}}$, $\phi_{\tilde{b}}$ enter
in the couplings when the weak eigenstates $\tilde{q}_{L}, \tilde{q}_{R}$
above are transformed into the mass eigenstates $\tilde{q}_{1}, \tilde{q}_{2}$.
In this paper we denote the vertices in squark mass eigenstate basis as
$$
\bar{t}-\tilde{b}_{i}-\tilde{\chi}^{+}_{j}:~~
V_{t\tilde{b}_{i}\tilde{\chi}^{+}_{j}}^{(1)}P_L+
V_{t\tilde{b}_{i}\tilde{\chi}^{+}_{j}}^{(2)}P_R,
\eqno{(2.8.1)}
$$
$$
t-\bar{\tilde{b}}_{i}-\bar{\tilde{\chi}}^{+}_{j}:~~
-V_{t\tilde{b}_{i}\tilde{\chi}^{+}_{j}}^{(2)\ast}P_L-
V_{t\tilde{b}_{i}\tilde{\chi}^{+}_{j}}^{(1)\ast}P_R,
\eqno{(2.8.2)}
$$
$$
\bar{t}-\tilde{t}_{i}-\tilde{\chi}^{0}_{j}:~~
V_{t\tilde{t}_{i}\tilde{\chi}^{0}_{j}}^{(1)}P_L+
V_{t\tilde{t}_{i}\tilde{\chi}^{0}_{j}}^{(2)}P_R,
\eqno{(2.8.3)}
$$
$$
t-\bar{\tilde{t}}_{i}-\bar{\tilde{\chi}}^{0}_{j}:~~
-V_{t\tilde{t}_{i}\tilde{\chi}^{0}_{j}}^{(2)\ast}P_L-
V_{t\tilde{t}_{i}\tilde{\chi}^{0}_{j}}^{(1)\ast}P_R,
\eqno{(2.8.4)}
$$
respectively, where $P_{L,R}=\frac{1}{2} (1\mp\gamma_5)$ and the explicit
expressions of the notations defined in Eqs.$(2.8.1)\sim(2.8.4)$ are listed
in Eqs.$(A.4)\sim(A.11)$ in Appendix A.
\par
  For the Feynman rules of the Higgs-quark-quark, Higgs-squark-squark,
Higgs-chargino-chargino and Z$(\gamma)$-chargino-chargino, one can refer
to Ref.\cite{a6}.
The couplings of $Higgs(B)-\tilde{\chi}^{+}_{k}-\tilde{\chi}^{+}_{k}$ have
a general form as
\begin{eqnarray*}
V_{B\tilde{\chi}^{+}_{k}\tilde{\chi}^{+}_{k}}=
V_{B\tilde{\chi}^{+}_{k}\tilde{\chi}^{+}_{k}}^{s}+
V_{B\tilde{\chi}^{+}_{k}\tilde{\chi}^{+}_{k}}^{ps} \gamma_5~~
(B=h^0, H^0, A^0, G^0),~~~~~~~~~~~~~~~~~~~~~~~~~~~~~~(2.9)
\end{eqnarray*}
The notations defined above which appear in the form factors, are
explicitly expressed in Eqs.$(A.12)\sim (A.15)$ of Appendix A.
\par
For Higgs-quark-quark and Higgs-squark-squark couplings, we denote
$$
H^0-t-t:~~
V_{H^{0}tt} = \frac{-i g m_t \sin{\alpha}}{2 m_W \sin{\beta}},
\eqno{(2.10.1)}
$$
$$
h^0-t-t:~~
V_{h^{0}tt} = \frac{-i g m_t \cos{\alpha}}{2 m_W \sin{\beta}},
\eqno{(2.10.2)}
$$
$$
A^0-t-t:~~
V_{A^{0}tt}\gamma_5 = \frac{-g m_t \cot{\beta}}{2 m_W}\gamma_5,
\eqno{(2.10.3)}
$$
$$
G^0-t-t:~~
V_{G^{0}tt}\gamma_5 = \frac{-g m_t}{2 m_W}\gamma_5,
\eqno{(2.10.4)}
$$
\par
The couplings of $H^0(h^0)-\tilde{q}_{i}-\tilde{q}_{i}~~(i=1,2,q=t,b)$ are
\begin{eqnarray*}
V_{H^0\tilde{t}_{1}\tilde{t}_{1}} &=&
  \frac{-i g m_Z \cos{(\alpha+\beta)}}{\cos{\theta_W}} \left[
   (\frac{1}{2} - \frac{2}{3} \sin^2 \theta_W) \cos^2\theta_{\tilde{t}} +
    \frac{2}{3} \sin^2 \theta_W \sin^2\theta_{\tilde{t}} \right] \\
&& - \frac{i g m_t^2 \sin{\alpha}}{m_W \sin\beta}
   + \frac{i g m_t}{2 m_W \sin{\beta}} (A_t \sin{\alpha} + \mu \cos{\alpha})
     \sin{\theta_{\tilde{t}}} \cos{\theta_{\tilde{t}}},~~~~(2.11.1)
\end{eqnarray*}
\begin{eqnarray*}
V_{H^0\tilde{t}_{2}\tilde{t}_{2}} &=&
  \frac{-i g m_Z \cos(\alpha+\beta)}{\cos\theta_W} \left[
 (\frac{1}{2} - \frac{2}{3} \sin^2 \theta_W) \sin^2\theta_{\tilde{t}} +
  \frac{2}{3} \sin^2 \theta_W \cos^2\theta_{\tilde{t}} \right] \\
&& - \frac{i g m_t^2 \sin\alpha}{m_W \sin\beta}
   - \frac{i g m_t}{2 m_W \sin\beta} (A_t \sin\alpha + \mu \cos\alpha)
     \sin\theta_{\tilde{t}} \cos\theta_{\tilde{t}},~~~~(2.11.2)
\end{eqnarray*}
\begin{eqnarray*}
V_{H^0\tilde{b}_{1}\tilde{b}_{1}} &=&
  \frac{i g m_Z \cos(\alpha+\beta)}{\cos\theta_W} \left[
 (\frac{1}{2} - \frac{1}{3} \sin^2 \theta_W) \cos^2\theta_{\tilde{b}} +
  \frac{1}{3} \sin^2 \theta_W \sin^2\theta_{\tilde{b}} \right] \\
&& - \frac{i g m_b^2 \cos\alpha}{m_W \cos\beta}
   + \frac{i g m_b}{2 m_W \cos\beta} (A_b \cos\alpha + \mu \sin\alpha)
     \sin\theta_{\tilde{b}} \cos\theta_{\tilde{b}},~~~~(2.11.3)
\end{eqnarray*}
\begin{eqnarray*}
V_{H^0\tilde{b}_{2}\tilde{b}_{2}} &=&
  \frac{i g m_Z \cos(\alpha+\beta)}{\cos\theta_W} \left[
 (\frac{1}{2} - \frac{1}{3} \sin^2 \theta_W) \sin^2\theta_{\tilde{b}} +
  \frac{1}{3} \sin^2 \theta_W \cos^2\theta_{\tilde{b}} \right] \\
&& - \frac{i g m_b^2 \cos\alpha}{m_W \cos\beta}
   - \frac{i g m_b}{2 m_W \cos\beta} (A_b \cos\alpha + \mu \sin\alpha)
     \sin\theta_{\tilde{b}} \cos\theta_{\tilde{b}},~~~~(2.11.4)
\end{eqnarray*}
\begin{eqnarray*}
V_{h^0\tilde{t}_{1}\tilde{t}_{1}} &=&
  \frac{i g m_Z \sin(\alpha+\beta)}{\cos\theta_W} \left[
 (\frac{1}{2} - \frac{2}{3} \sin^2 \theta_W) \cos^2\theta_{\tilde{t}} +
  \frac{2}{3} \sin^2 \theta_W \sin^2\theta_{\tilde{t}} \right] \\
&& - \frac{i g m_t^2 \cos\alpha}{m_W \sin\beta}
   + \frac{i g m_t}{2 m_W \sin\beta} (A_t \cos\alpha - \mu \sin\alpha)
     \sin\theta_{\tilde{t}} \cos\theta_{\tilde{t}},~~~~(2.11.5)
\end{eqnarray*}
\begin{eqnarray*}
V_{h^0\tilde{t}_{2}\tilde{t}_{2}} &=&
  \frac{i g m_Z \sin(\alpha+\beta)}{\cos\theta_W} \left[
 (\frac{1}{2} - \frac{2}{3} \sin^2 \theta_W) \sin^2\theta_{\tilde{t}} +
  \frac{2}{3} \sin^2 \theta_W \cos^2\theta_{\tilde{t}} \right] \\
&& - \frac{i g m_t^2 \cos\alpha}{m_W \sin\beta}
   - \frac{i g m_t}{2 m_W \sin\beta} (A_t \cos\alpha - \mu \sin\alpha)
     \sin\theta_{\tilde{t}} \cos\theta_{\tilde{t}},~~~~(2.11.6)
\end{eqnarray*}
\begin{eqnarray*}
V_{h^0\tilde{b}_{1}\tilde{b}_{1}} &=&
 \frac{-i g m_Z \sin(\alpha+\beta)}{\cos\theta_W} \left[
(\frac{1}{2} - \frac{1}{3} \sin^2 \theta_W) \cos^2\theta_{\tilde{b}} +
 \frac{1}{3} \sin^2 \theta_W \sin^2\theta_{\tilde{b}} \right] \\
&& + \frac{i g m_b^2 \sin\alpha}{m_W \cos\beta}
   - \frac{i g m_b}{2 m_W \cos\beta} (A_b \sin\alpha - \mu \cos\alpha)
     \sin\theta_{\tilde{b}} \cos\theta_{\tilde{b}},~~~~(2.11.7)
\end{eqnarray*}
\begin{eqnarray*}
V_{h^0\tilde{b}_{2}\tilde{b}_{2}} &=&
  \frac{-i g m_Z \sin(\alpha+\beta)}{\cos\theta_W} \left[
 (\frac{1}{2} - \frac{1}{3} \sin^2 \theta_W) \sin^2\theta_{\tilde{b}} +
  \frac{1}{3} \sin^2 \theta_W \cos^2\theta_{\tilde{b}} \right] \\
&& + \frac{i g m_b^2 \sin\alpha}{m_W \cos\beta}
   + \frac{i g m_b}{2 m_W \cos\beta} (A_b \sin\alpha - \mu \cos\alpha)
     \sin\theta_{\tilde{b}} \cos\theta_{\tilde{b}},~~~~(2.11.8)
\end{eqnarray*}
respectively.

\vskip 5mm
\noindent{\large\bf III. Calculations}

\par
In the calculation, we take the t'Hooft gauge and adopt the dimensional
reduction (DR) scheme \cite{Copper}, which is commonly used in the calculations
of the MSSM radiative corrections as it preserves supersymmetry at
least at one-loop order, to eliminate the ultraviolet divergences in the
virtual loop corrections. We choose the on-mass-shell (OMS) scheme
\cite{se} for doing renormalization.

\par
\begin{center} {\bf 3.1 The tree-level formulae and notations.}\end{center}

\par
In the process of top-pair production via photon-photon collision,
the Mandelstam variables $\hat{s}$, $\hat{t}$ and $\hat{u}$ are defined
as $\hat{s}=(p_{1}+p_{2})^2,~~\hat{t}=(p_1-p_3)^2,~~\hat{u}=(p_1-p_4)^2$.
The corresponding Lorentz invariant matrix element at the lowest order for the
reaction $\gamma \gamma \rightarrow t \bar{t}$ is written as
$$
{\cal M}_{0}={\cal M}_{\hat{t}}+{\cal M}_{\hat{u}},
\eqno{(3.1.1)}
$$
where
$$
{\cal M}_{\hat{t}}=\left[ \bar{u}(p_3)(-i e \gamma_{\mu})
\frac{i}{\hat{\rlap/t}-m_t} (-i e \gamma_{\nu}) v(p_4)
\epsilon^{\mu}(p1) \epsilon^{\nu}(p2) \right],
\eqno{(3.1.2)}
$$
$$
{\cal M}_{\hat{u}}=\left[ \bar{u}(p_3)(-i e \gamma_{\nu})
\frac{i}{\hat{\rlap/u}-m_t} (-i e \gamma_{\mu}) v(p_4)
\epsilon^{\nu}(p2) \epsilon^{\mu}(p1) \right].
\eqno{(3.1.3)}
$$
\par
The corresponding differential cross section is obtained by
$$
\frac{d \hat{\sigma}_{0}(\hat{t},\hat{s})} {d \hat{t}}=
   \frac{N_c}{16\pi^2\hat{s}}\bar{\sum_{spins}}|{\cal M}_{0}|^2,
\eqno{(3.1.4)}
$$
where the summation with a bar over head means to sum up the spins of final
states and average the spins of initial photons. After integration over
$\hat{t}$, the total Born cross section with unpolarized incoming photons is
worked out as
$$
\hat{\sigma}_{0}(\hat{s})=\frac{32\pi \alpha^2}{27 \hat{s}} \left[
           2\hat{\beta}(\hat{\beta}^2-2) +(3-\hat{\beta}^4)
           \ln{ \frac{1+\hat{\beta}}{1-\hat{\beta}} }\right].
\eqno{(3.1.5)}
$$
where the kinematic factor is defined as
$$
\hat{\beta}=\sqrt{1-4 m_t^2/\hat{s}}.
\eqno{(3.1.6)}
$$
\par
The total cross section including the leading
one-loop corrections in the frame of the MSSM is
$$
\hat{\sigma} = \hat{\sigma}_{0} + \delta \hat{\sigma}^{1-loop},
\eqno{(3.1.7)}
$$
where $\delta \hat{\sigma}^{1-loop}$ represents the interference term
between tree-level and one-loop correction amplitudes.

\par
\begin{center} {\bf 3.2 Self-energies.}
\end{center}

\par
The top quark wave function corrections $\delta Z_{tt}$'s are determined
in terms of the one-particle irreducible two-point function $i\Gamma(p^2)$
for top quarks in the DR mass basis. It should be written as\cite{a12}:
$$
\begin{array} {lll}
\Gamma_{tt}(p^2) &=&
         (\rlap/p-m_{t}) +  \left [ \rlap/p P_{L}
    \Sigma_{tt}^{L}(p^2)
   + \rlap/p P_{R} \Sigma_{tt}^{R}(p^2)
   + P_{L} \Sigma_{tt}^{S,L}(p^2)
   + P_{R} \Sigma_{tt}^{S,R}(p^2) \right].
\end{array}
\eqno{(3.2.1)}
$$
\par
  With the Feynman rules of the interactions of top-sbottom-chargino
and top-stop-neutralino, the corresponding unrenormalized chargino
self-energies read (see Fig.1(f))
\begin{eqnarray*}
\Sigma^{S,L}_{tt}(p^2) = \frac{1}{16 \pi^2} \sum_{j=1,2} && \left(
  \sum_{i=1,4} m_{\tilde{\chi}_i^0}
  V_{t\tilde{t}_j\tilde{\chi}_i^0}^{(1)}
  V_{t\tilde{t}_j\tilde{\chi}_i^0}^{(2)\ast}
  B_0[-p, m_{\tilde{\chi}_i^0}, m_{\tilde{t}_j}] \right. \\
&& + \sum_{i=1,2} m_{\tilde{\chi}_i^+} \left.
  V_{t\tilde{b}_j\tilde{\chi}_i^+}^{(1)}
  V_{t\tilde{b}_j\tilde{\chi}_i^+}^{(2)\ast}
  B_0[-p, m_{\tilde{\chi}_i^+}, m_{\tilde{b}_j}] \right),
~~~~~~~~~~~~~~~~(3.2.2)
\end{eqnarray*}
\begin{eqnarray*}
\Sigma^{S,R}_{tt}(p^2) = \frac{1}{16 \pi^2} \sum_{j=1,2} && \left(
  \sum_{i=1,4} m_{\tilde{\chi}_i^0}
  V_{t\tilde{t}_j\tilde{\chi}_i^0}^{(2)}
  V_{t\tilde{t}_j\tilde{\chi}_i^0}^{(1)\ast}
  B_0[-p, m_{\tilde{\chi}_i^0}, m_{\tilde{t}_j}] \right. \\
&& + \sum_{i=1,2} m_{\tilde{\chi}_i^+} \left.
  V_{t\tilde{b}_j\tilde{\chi}_i^+}^{(2)}
  V_{t\tilde{b}_j\tilde{\chi}_i^+}^{(1)\ast}
  B_0[-p, m_{\tilde{\chi}_i^+}, m_{\tilde{b}_j}] \right),
~~~~~~~~~~~~~~~(3.2.3)
\end{eqnarray*}
$$
\Sigma^{L}_{tt}(p^2) = -\frac{1}{16 \pi^2} \sum_{j=1,2} \left(
  \sum_{i=1,4}
  |V_{t\tilde{t}_j\tilde{\chi}_i^0}^{(2)}|^2
  B_1[-p, m_{\tilde{\chi}_i^0}, m_{\tilde{t}_j}] +
  \sum_{i=1,2}
  |V_{t\tilde{b}_j\tilde{\chi}_i^+}^{(2)}|^2
  B_1[-p, m_{\tilde{\chi}_i^+}, m_{\tilde{b}_j}] \right),
\eqno{(3.2.4)}
$$
$$
\Sigma^{R}_{tt}(p^2) =-\frac{1}{16 \pi^2} \sum_{j=1,2} \left(
  \sum_{i=1,4}
  |V_{t\tilde{t}_j\tilde{\chi}_i^0}^{(1)}|^2
  B_1[-p, m_{\tilde{\chi}_i^0}, m_{\tilde{t}_j}] +
  \sum_{i=1,2}
  |V_{t\tilde{b}_j\tilde{\chi}_i^+}^{(1)}|^2
  B_1[-p, m_{\tilde{\chi}_i^+}, m_{\tilde{b}_j}] \right).
\eqno{(3.2.5)}
$$
\par
Imposing the on-shell renormalization conditions given in Ref.\cite{se}
\cite{a12}, one can obtain the renormalization constants for the
renormalized top quark self-energies as\cite{a5}:
$$
\delta \Sigma_{tt}(p^2) = C_{L} \rlap/p P_{L} +
   C_{R} \rlap/p P_{R} - C^{-}_{S} P_{L} - C^{+}_{S} P_{R},
\eqno{(3.2.6)}
$$
\par
The $\gamma\gamma$ and $\gamma Z^0$ self-energies with only quark and squark
one-loops were presented in reference\cite{MZhou}. We can see that the
self-energies of $\gamma\gamma$ and $\gamma Z^0$ have no contribution to
the relevant counterterms of the $\gamma tt$ vertex. The renormalization
constant for the $\Gamma^{\mu}_{\gamma tt}$ vertex is written in the form of
$$
 \delta \Gamma^{\mu}_{\gamma tt} =
   -i e \gamma^{\mu}[ C^{L} P_{L} + C^{R} P_{R} ].
\eqno{(3.2.7)}
$$
where
$$
\begin{array} {lll}
C_L &=& \frac{1}{2} (\delta Z^L_{tt} +
   \delta Z^{L\dag}_{tt}), \\
C_R &=& \frac{1}{2} (\delta Z^R_{tt} +
   \delta Z^{R\dag}_{tt}), \\
C^{-}_{S} &=& \frac{m_t}{2}
(\delta Z^L_{tt} +
   \delta Z^{R\dag}_{tt}) +
   \delta m_t, \\
C^{+}_{S} &=& \frac{m_t}{2}
   (\delta Z^R_{tt} +
   \delta Z^{L\dag}_{tt}) +
   \delta m_t.
\end{array}
\eqno{(3.2.8)}
$$
\begin{eqnarray*}
\delta m_t &=& \frac{1}{2} \tilde{Re}
  \left [ m_t \Sigma^{L}_{tt} (m_t^2) + m_t
  \Sigma^{R}_{tt}(m_t^2) +
  \Sigma^{S,L}_{tt}(m_t^2) +
  \Sigma^{S,R}_{tt}(m_t^2)
  \right ],~~~~~(3.2.9)
\end{eqnarray*}
\begin{eqnarray*}
\delta Z^{L}_{tt} &=&
- \tilde{Re}\Sigma^{L}_{tt} (m_t^2) - \frac{1}{m_t} \tilde{Re}
  \left [ \Sigma^{S,R}_{tt} (m_t^2)
  - \Sigma^{S,L}_{tt} (m_t^2)
  \right ] \\
&-& m_t \frac{\partial}{\partial p^2} \tilde{Re}
  \left \{ m_t \Sigma^{L}_{tt}(p^2)
  + m_t \Sigma^{R}_{tt}(p^2)
  \right.\\
&+& \left.  \Sigma^{S,L}_{tt}(p^2) +
     \Sigma^{S,R}_{tt}(p^2) \right\} |_
     {p^2=m_t^2},
~~~~~~(3.2.10)
\end{eqnarray*}
\begin{eqnarray*}
\delta Z^{R}_{tt} &=& -\tilde{Re}\Sigma^{R}_{tt} (m_t^2) -
  m_t \frac{\partial} {\partial p^2} \tilde{Re}
  \left\{ m_t \Sigma^{L}_{tt}(p^2) +
  m_t \Sigma^{R}_{tt}(p^2) \right. \\
&+& \left. \Sigma^{S,L}_{tt}(p^2) +
    \Sigma^{S,R}_{tt}(p^2) \right \} |_
    {p^2=m_t^2}, ~~~~~~~(3.2.11)
\end{eqnarray*}
where $\tilde{Re}$ takes the real part of the loop integrals. It ensures
reality of the renormalized lagrangian.

\par
\begin{center} {\bf 3.3 Renormalized one-loop corrections.} \end{center}
\par
    The renormalized one-loop matrix element involves the contributions from
all the self-energy, vertex, box, triangle and quartic interaction
one-loop diagrams and their relevant counterterms.
The Feynman diagrams for the process (2.1) are depicted in Fig.1,
where (a) is for the tree-level and (b) $\sim$ (f) are EW-like one-loop
diagrams contributing to the cross section in the frame of the MSSM.
Specifically, Fig.1(b.1 $\sim$ 4) are the vertex diagrams, Fig.1(c.1
$\sim$ 3) are the box diagrams, Fig.1(d.1 $\sim$ 2) are the quartic
interactions, Fig.1(e.1 $\sim$ 2) are the triangle sectors, and Fig.1(f)
is the self-energy diagram.  In below, we denote them
by the upper indexes of $v$, $b$, $q$, $tr$ and $self$, respectively.
The relevant Feynman rules are shown in section II\cite{a6}.
In the calculation, some of the s-channel Feynman diagrams involving quark
loops with the exchanging of $\gamma$ or $Z^0$ boson in Fig.1(e.2) can be
neglected, as the consequence of Furry theorem. It is because
that the Furry theorem forbids the production of the spin-one components of
the $Z^{0}$ and $\gamma$, and the contribution from the spin-zero component
of the $Z^{0}$ vector boson coupling with a pair of chargino is very small
and neglectable. The calculation also shows the $\gamma$ and $Z^0$ exchanging
s-channel diagrams in Fig.1(d.2) and Fig.1(e.1) with a squark loop have no
contribution to the cross section, in which the contribution from each of the
$\gamma$ and $Z^0$ exchanging s-channel diagrams in Fig.1(e.1) is canceled out
by the corresponding one with exchanging incoming photons.
Including all the diagrams appearing in Fig.1, the renormalized matrix
elements for $t\bar{t}$ pair production in $\gamma\gamma$ collision is
written as
\begin{eqnarray*}
\delta {\cal M}_{1-loop} & = &
    {\cal M}^{v}+{\cal M}^{b}+{\cal M}^{q}+{\cal M}^{tr}+{\cal M}^{self}\\
&=& {\cal M}^{v,\hat{t}}+{\cal M}^{v,\hat{u}}+{\cal M}^{b,\hat{t}}+
    {\cal M}^{b,\hat{u}}+{\cal M}^{q}+{\cal M}^{tr,\hat{t}}+
    {\cal M}^{tr,\hat{u}}+{\cal M}^{self,\hat{t}}+{\cal M}^{self,\hat{u}}\\
&=& \epsilon^{\mu}(p_3)\epsilon^{\nu}(p_4) \bar{u}(p_1) \left\{
   f_{1} \gamma_{\mu}\gamma_{\nu} + f_{2} \gamma_{\nu}\gamma_{\mu} +
   f_{3} \gamma_{\mu}p_{1\nu} + f_{4} \gamma_{\mu}p_{2\nu} \right. \\
&+& \left. f_{5} \gamma_{\nu}p_{1\mu} + f_{6} \gamma_{\nu}p_{2\mu} +
   f_{7}  p_{1\mu}p_{1\nu} + f_{8}  p_{1\mu}p_{2\nu} +
   f_{9}  p_{1\nu}p_{2\mu} \right. \\
&+& \left. f_{10} p_{2\mu}p_{2\nu} +
   f_{11} \rlap/{p}_{3} \gamma_{\mu} \gamma_{\nu} +
   f_{12} \rlap/{p}_{3} \gamma_{\nu} \gamma_{\mu} +
   f_{13} \rlap/{p}_{3} \gamma_{\mu} p_{1\nu} +
   f_{14} \rlap/{p}_{3} \gamma_{\mu} p_{2\nu} \right.  \\
&+& \left. f_{15} \rlap/{p}_{3} \gamma_{\nu} p_{1\mu} +
   f_{16} \rlap/{p}_{3} \gamma_{\nu} p_{2\mu} +
   f_{17} \rlap/{p}_{3} p_{1\mu} p_{1\nu} +
   f_{18} \rlap/{p}_{3} p_{1\mu} p_{2\nu} \right.  \\
&+& \left. f_{19} \rlap/{p}_{3} p_{1\nu} p_{2\mu} +
   f_{20} \rlap/{p}_{3} p_{2\mu} p_{2\nu} +
   f_{21} \gamma_5 \epsilon_{\mu\nu\alpha\beta} p_{1}^{\alpha} p_{3}^{\beta}
   \right.  \\
&+& \left. f_{22} \gamma_5 \epsilon_{\mu\nu\alpha\beta} p_{2}^{\alpha}
   p_{3}^{\beta} \right\} v(p_2),~~~~~~~~~~~~~~~~(3.3.1)
\end{eqnarray*}
with form factors
$$
f_i = f_{i}^{v}+f_{i}^{b}+f_{i}^{q}+f_{i}^{tr}+f_{i}^{self}~~~
(i=1 \sim 22),
\eqno{(3.3.2)}
$$
\par
Here we have divided each matrix element ${\cal M}^{v}$, ${\cal M}^{b}$,
${\cal M}^{tr}$ and ${\cal M}^{self}$ into t-channel and u-channel parts.
For each of the corresponding form factor we have
$$
f_{i}^{k}=f_{i}^{k,\hat{t}}+f_{i}^{k,\hat{u}},~~~~
(k=v,b,tr,self, ~~i=1 \sim 22),
\eqno{(3.3.3)}
$$
\par
The vertex, box and triangle diagrams with exchanging photons(i.e.,
u-channel) are not shown in Fig.1.  The amplitude parts from the u-channel
vertex, box and quartic interaction corrections can be obtained from the
t-channel's by doing exchanges as below:
\begin{eqnarray*}
{\cal M}^{j,\hat{u}}={\cal M}^{j,\hat{t}}(t \rightarrow u,
   p_3 \leftrightarrow p_4, \mu \leftrightarrow \nu),~~(j=v,b,tr,s)
   ~~~~~~~~~~~~~~~~~~~~~~~~~~~~~~~~(3.3.4)
\end{eqnarray*}
\par
Then we list only the explicit t-channel form factors in Appendix B.
Now we can obtain the one-loop corrections to the cross section from the
chargino and neutralino sectors for this subprocess in unpolarized photon
collisions.
\begin{eqnarray*}
 \delta \hat{\sigma}^{1-loop}(\hat{s}) &=& \frac{N_c}{16 \pi \hat{s}^2}
             \int_{\hat{t}^{-}}^{\hat{t}^{+}} d\hat{t}~
        2 Re {\bar{\sum\limits_{spins}^{}}} \left( {\cal M}_{0}^{\dag} \cdot
        \delta {\cal M}_{1-loop} \right),~~~~~~~(3.3.5)
\end{eqnarray*}
where $\hat{t}^\pm=(m_t^2-\frac{1}{2}\hat{s}) \pm\frac{1}{2}\hat{s} \beta$.
The cross section of the top-pair production via photon-photon fusion
at the $e^{+}e^{-}$ linear collider, can be obtained by folding the cross
section of the subprocess $\hat{\sigma} (\gamma\gamma \rightarrow t\bar{t}$
with the photon luminosity\cite{sd,sh}\cite{sd,si}.

\vskip 5mm
\noindent{\large\bf IV. Numerical Results and Discussions}

\par
The SUSY EW-like corrections to top-pair production process are
strongly related to the fundamental MSSM parameters through the electroweak
couplings involving top-quark, squark and chargino (neutralino), i.e.
$V_{t\tilde{b}\tilde{\chi}^{+}}$ and $V_{t\tilde{t}\tilde{\chi}^{0}}$
as expressed in Eqs.$(2.8.1 \sim 4)$. For our numerical calculation
of squark sector, we take $M_{Q}$, $\theta_{\tilde{t}}$ and $\theta_{\tilde{b}}$
as input parameters, and we set $\theta_{\tilde{b}}=0$, and $\theta_{\tilde{t}}$
approaches $\frac{\pi}{4}$, so that the masses of top squark pair split
remarkably, while the split of the sbottom masses is minimized. From the
Eq.(2.7) and relevant expressions, we can see that the parameter $M_{Q}$
is strongly related to the masses of top and bottom squarks, therefore it
would affect the MSSM correction quantitatively in some regions of the
parameter space.
\par
As stated in Section II, the correction should also depend on the
fundamental MSSM parameters $\tan{\beta}$, $M_{SU(2)}$ and $\mu$ through
gaugino and higgsino couplings. Note that these parameters take parts in
the EW-like corrections not only through the chargino and
neutralino mass spectra, but also through the couplings including their
transformation matrices $U$, $V$ and $N$.
\par
We take some of the general constants as: $m_t=175~GeV$, $m_Z=91.187~GeV$,
$m_b=4.5~GeV$, $\sin^2{\theta_{W}}=0.2315$, and $\alpha = 1/128$. And we
adopt the following set of input parameters by default, in case that the
parameter is not set as the independent variable of the figure and no special
declaration has been presented on them:
$$
      \sqrt{\hat{s}}=500~or~1000~GeV,~~~ \tan\beta=4~or~40,
$$
$$
      M_{Q}=M_{SU(2)}=\mu=200~GeV,
$$
$$
      \theta_{\tilde{t}}=44.325^{\circ},~~~ \theta_{\tilde{b}}=0.
\eqno{(4.1)}
$$
\par
We use the analytical formulae for the masses of the MSSM Higgs bosons
(including two-loop leading-log corrections and squark mixing effects)
given in reference \cite{Esp1}\cite{Esp}.
$$
m_{h^0,H^0}^2 = \frac{1}{2}\left[TrM^2 \mp \sqrt{(TrM^2)^2-4 det M^2}\right],
\eqno{(4.2.1)}
$$
where
$$
Tr M^2=M_{11}^2+M_{22}^2, \hskip 5mm
det M^2 = M_{11}^2 M_{22}^2-(M_{12}^{2})^2,
\eqno{(4.2.2)}
$$
with
$$
M_{12}^2=2v^2\left[ sin\beta cos\beta(\lambda_3+\lambda_4)+
         \lambda_6 cos^2\beta+\lambda_7 sin^2\beta\right]-m_{A^0}^2
         sin\beta cos\beta
$$
$$
M_{11}^2=2v^2\left[ \lambda_1 cos^2\beta +2 \lambda_6 cos\beta sin\beta +
         \lambda_5 sin^2\beta \right]+m_{A^0}^2 sin^2\beta
$$
$$
M_{22}^2=2v^2\left[ \lambda_2 sin^2\beta +2 \lambda_7 cos\beta sin\beta +
         \lambda_5 cos^2\beta\right]+m_{A^0}^2 cos^2\beta,
\eqno{(4.2.3)}
$$
where $v=174.1~GeV$. The mixing angle $\alpha$ is determined by
$$
sin 2 \alpha = \frac{2 M_{12}^2}{\sqrt{(Tr M^2)^2-4 det M^2}},
\eqno{(4.2.4)}
$$
One can find the explicit expressions of $\lambda_{i}(i=1,...,7)$ in
reference\cite{Esp1}. In this work, we take $m_{A^0}=150~GeV$.
\par
Our numerical results are presented in figures.  In Fig.2(a) and Fig.2(b),
the correction $\Delta \sigma$ and the relative correction $\delta=\frac{\Delta
\sigma}{\sigma_0}$ of the process (2.1) depending on the c.m.s. energy
$\sqrt{\hat{s}}$ are plotted, respectively.  From our analyses, we expect
that for the curves of $\tan\beta=4$ in Fig.2(a) and Fig.2(b), there should
be some spikes or turning points at $\sqrt{\hat{s}} \sim 2 m_{\tilde{b}_{1}}=
403~GeV$, $2 m_{\tilde{b}_{2}}=415~GeV$, $2 m_{\tilde{\chi}^{+}_{2}}=546~GeV$,
and $2 m_{\tilde{t}_{2}}=628~GeV$ due to the resonance effects.
But we see in both figures that the first two resonance points merge each
other in the curves of $\tan{\beta}=4$ because they are too near.
For $\tan\beta=40$, there are only two obvious resonance points can be seen
on the curve in figure 2(a) in the vicinities of $\sqrt{\hat{s}} \sim
2 m_{\tilde{b}_{1}} \sim 2 m_{\tilde{b}_{2}} \sim 410~GeV$ and
$\sqrt{\hat{s}}=2 m_{\tilde{\chi}^{+}_{2}}=531~ GeV$.
While only one obvious resonance peak can be seen on the curve of
$\tan\beta=40$ at $\sqrt{\hat{s}}=2 m_{\tilde{\chi}^{+}_{2}}=531~ GeV$
in Fig.2(b).
\par
      In Fig.3 $\sim$ 6 we depicted the dependences of the relative radiative
correction on the fundamental supersymmetric input parameters $M_Q$,
$M_{SU(2)}$ and $|\mu|$, respectively. In each figure we take four data sets
for discussion: (1) $\tan\beta=4,~\sqrt{\hat{s}}=500~GeV$; (2) $\tan\beta=40,~
\sqrt{\hat{s}}=500~GeV$; (3) $\tan\beta=4,~\sqrt{\hat{s}}=1~TeV$; (4)
$\tan\beta=40,~ \sqrt{\hat{s}}=1~TeV$. From Fig.3 one can see the absolute
value of the relative correction becomes generally larger when $\sqrt{\hat{s}}$
goes higher. The same feature is also shown in Fig.2.  Fig.3 present that the
absolute value of the relative correction goes down to a smaller constant
with $M_Q$ increasing. Since $M_Q$ is related to the masses of squarks
$\tilde{t}_{1,2}$ and $\tilde{b}_{1,2}$ as stated in Eq.(2.7), it can be
easily understood as the feature of the decoupling effect. We can conclude
that the smaller $M_Q$ is, the more significant the correction can be. We can
read from Fig.3 that the relative correction can reach $- 2\%$ when $M_Q$ is
about $150~GeV$, and for large $M_Q$ with same $\sqrt{\hat{s}}$ the parameter
$\tan{\beta}$ tends to make little difference on the relative correction. In
Fig.4, the corrections as the function of $M_{SU(2)}$ are plotted with the
four data sets.  The grooves around $230~GeV$ on the two curves for $\sqrt
{\hat{s}}=500~GeV$ and the small heaves in the vicinity of $450~GeV$ on the
two curves of $\sqrt{\hat{s}}=1~TeV$, are all because of the resonance
effect: $\sqrt{\hat{s}} \sim 2 m_{\tilde{\chi}^{+}_{2}}$.  When $M_{SU(2)}$
is large, all of the four curves become very plain because of the decoupling
effect. In Fig.5, there are two peaks at the position about $\mu \sim 470~
GeV$ on the curves of $\sqrt{\hat{s}}=1~TeV$ due to the resonance effect,
but the resonance effects around the region $\mu \sim 230~GeV$ on the
curves with $\sqrt{\hat{s}}= 500~GeV$ are not clear. And we can see that the
correction is no longer sensitive to $\tan{\beta}$ and $\mu$ when $\mu$ gets
larger than $600~GeV$. This is because the parameter $\tan{\beta}$ and $\mu$
are not only related to the masses of sparticles, but also involved in some
vertices which are concerned in our calculation. Both Fig.4 and Fig.5 show
that the higher the c.m.s. energy $\sqrt{\hat{s}}$ is, the larger the relative
corrections to subprocess are.
\par
Fig.6 shows the cross section of the parent process $e^+ e^- \rightarrow
\gamma \gamma \rightarrow t \bar{t}$ including one-loop EW-like corrections
as the function of c.m.s energy of incoming electron-positron pair.
In Fig.7, the relative corrections for $\tan{\beta}=4$ and $\tan{\beta}=40$
are plotted, respectively. It is clear that the absolute value of the relative
correction becomes larger with the increasing of the $e^+ e^-$ c.m.s energy.
The reduction of the cross section of the parent process due to the one-loop
EW-like correction can approach to one percent.

\vskip 10mm
\noindent
{\Large{\bf V.Summary}}
\vskip 5mm
In this work we have studied the complete one-loop radiative corrections
from the gaugino-Higgsino-sector in the process $\gamma \gamma \rightarrow
t \bar{t}$ in the frame of the MSSM at the NLC. This process
has great importance at the future NLC operating in photon-photon
collision mode. From the numerical calculation with several typical sets of
input parameters, we find that the EW-like corrections from the
chargino/neutralino sector can be a few percent
for subprocess and can approach to one percent for the parent process. These
corrections are smaller than the QCD corrections, but are comparable to
the electroweak correction part from the Higgs sector in the MSSM. Therefore
the correction from chargino/neutralino sector is also significant and
unneglectable. We investigated also the dependences of
the corrections on the supersymmetric parameters. With the variation of the
parameters $M_Q$, $M_{SU(2)}$ and $|\mu|$, we can see some physical features,
such as the decoupling effects, threshold effects and the resonance effects,
where the relative correction can be significantly enhanced or diminished.
We conclude that the EW-like one-loop correction to Born cross section is
strongly dependent on the c.m.s. energy and the related MSSM parameters
in some cases. We find that the correction is not sensitive to $M_{SU(2)}$
(or $|\mu|$) when $M_{SU(2)}~>>~|\mu|$ (or $|\mu|~>>~M_{SU(2)}$). The
correction is weakly dependent on the ratio of the vacuum expectation values
$\tan{\beta}$, when $M_Q$ (or $|\mu|$) is large enough. But it is related to
the c.m. energy of the incoming photons obviously.\\

\vskip 5mm
\noindent{\large\bf Acknowledgement:}
These work was supported in part by the National Natural Science
Foundation of China(project numbers: 19675033, 19875049) and the
Youth Science Foundation of the University of Science and Technology
of China.

\vskip 5mm
\begin{center} {\Large Appendix}\end{center}
{\large A. Some expressions defined in Lagrangian.}
\par
In the lagrangian shown in eq.(2.6), we denote
$\tilde{q}_{L}$ and $\tilde{q}_{R}$ as the current eigenstates.
For the up-type scalar quarks, we have
$$
m^2_{\tilde{q}_{L}}=\tilde{M}^2_{Q} + m^2_{q} +
      m_{Z}^2 (\frac{1}{2}- Q_q s_{W}^2) \cos{2 \beta},$$
$$
m^2_{\tilde{q}_{R}}=\tilde{M}^2_{U} + m^2_{q} +
      Q_q m_{Z}^2 s_{W}^2 \cos{2 \beta},$$
$$
a_{q}=\mu \cot{\beta} + A_{q} \tilde{M}.
\eqno{(A.1)}
$$
\par
For the down-type scalar quarks,
$$m^2_{\tilde{q}_{L}}=\tilde{M}^2_{Q} + m^2_{q}
      - m_{Z}^2 (\frac{1}{2}+ Q_q s_{W}^2) \cos{2 \beta} ,$$
$$m^2_{\tilde{q}_{R}}=\tilde{M}^2_{D} + m^2_{q}
      + Q_q m_{Z}^2 s_{W}^2 \cos{2 \beta},$$
$$
a_{q}=\mu \tan{\beta} + A_{q} \tilde{M},
\eqno{(A.2)}
$$
where $Q_{q}=\frac{2}{3}$(for up-type), $-\frac{1}{3}$(for down-type) is the
charge of the scalar quark, $\tilde{M}^2_{Q}$, $\tilde{M}^2_{U}$ and
$\tilde{M}^2_{D}$ are the self-supersymmetry-breaking mass terms for
the left-handed and right-handed scalar quarks, and $s_W=\sin{\theta_W}$,
$c_W=\sin{\theta_W}$. As an assumption at Planck scale, We choose
$\tilde{M}_Q = \tilde{M}_U = \tilde{M}_D = \tilde{M}$.  Since CP effects are
not considered, the value $a_{q}$ is real. When $\tilde{q}_L$ and
$\tilde{q}_R$ are mixed, they give the mass eigenstates $\tilde{q}_1$
and $\tilde{q}_2$. The mass eigenstates $\tilde{q}_1$ and $\tilde{q}_2$
are expressed in terms of the current eigenstates $\tilde{q}_L$, $\tilde{q}_R$
as
read
$$
\tilde{q}_1=\tilde{q}_L \cos{\theta_q} - \tilde{q}_R \sin{\theta_q} ,
$$
$$
\tilde{q}_2= \tilde{q}_L \sin{\theta_q} + \tilde{q}_R \cos{\theta_q},
$$
with
$$
\tan{2 \theta_q}=\frac{2 a_{q} m_{q}}
                      {m^2_{\tilde{q}_{L}}-m^2_{\tilde{q}_{R}} }.
\eqno{(A.3)}
$$
\par
The explicit expressions for the notations used in Eqs.$(2.8.1)\sim(2.8.4)$
are listed as below
$$
V_{t\tilde{b}_{1}\tilde{\chi}^{+}_{j}}^{(1)}= \frac{i g m_t}
  {\sqrt{2}m_W \sin{\beta}} V_{j2}^{\ast} \cos{\theta_{\tilde{b}}},
\eqno{(A.4)}
$$
$$
V_{t\tilde{b}_{1}\tilde{\chi}^{+}_{j}}^{(2)}= -i g (U_{j1} \cos{\theta_{\tilde{b}}}
   + \frac{m_b}{\sqrt{2}m_W \cos{\beta}} U_{j2} \sin{\theta_{\tilde{b}}}),
\eqno{(A.5)}
$$
$$
V_{t\tilde{b}_{2}\tilde{\chi}^{+}_{j}}^{(1)}= \frac{i g m_t}
  {\sqrt{2}m_W \sin{\beta}} V_{j2}^{\ast} \sin{\theta_{\tilde{b}}},
\eqno{(A.6)}
$$
$$
V_{t\tilde{b}_{2}\tilde{\chi}^{+}_{j}}^{(2)}= -i g (U_{j1} \sin{\theta_{\tilde{b}}}
   - \frac{m_b}{\sqrt{2}m_W \cos{\beta}} U_{j2} \cos{\theta_{\tilde{b}}}),
\eqno{(A.7)}
$$
$$
V_{t\tilde{t}_{1}\tilde{\chi}^{0}_{j}}^{(1)}= -i g \sqrt{2} (\frac{m_t}{2 m_W
   \sin\beta} N_{j4}^{\ast} \cos{\theta_{\tilde{t}}} + \frac{2}{3} \tan\theta_W
   N_{j1}^{\ast} \sin{\theta_{\tilde{t}}}),
\eqno{(A.8)}
$$
$$
V_{t\tilde{t}_{1}\tilde{\chi}^{0}_{j}}^{(2)}= -i g \sqrt{2} ((\frac{1}{6}
   \tan\theta_W N_{j1}+\frac{1}{2} N_{j2}) \cos{\theta_{\tilde{t}}} -
   \frac{m_t}{2 m_W \sin\beta} N_{j4} \sin{\theta_{\tilde{t}}}),
\eqno{(A.9)}
$$
$$
V_{t\tilde{t}_{2}\tilde{\chi}^{0}_{j}}^{(1)}= -i g \sqrt{2} (\frac{m_t}{2 m_W
   \sin\beta} N_{j4}^{\ast} \sin{\theta_{\tilde{t}}} - \frac{2}{3} \tan\theta_W
   N_{j1}^{\ast} \cos{\theta_{\tilde{t}}}),
\eqno{(A.10)}
$$
$$
V_{t\tilde{t}_{2}\tilde{\chi}^{0}_{j}}^{(2)}= -i g \sqrt{2} ((\frac{1}{6}
   \tan\theta_W N_{j1}+\frac{1}{2} N_{j2}) \sin{\theta_{\tilde{t}}} +
   \frac{m_t}{2 m_W \sin\beta} N_{j4} \cos{\theta_{\tilde{t}}}),
\eqno{(A.11)}
$$
\par
The shorted notations defined in Eq.(2.9), are explicitly expressed below.
$$
V_{H^0\tilde{\chi}^{+}_{k}\tilde{\chi}^{+}_{k}}^{s} =
  \frac{-i g}{\sqrt{2}} \left[ \cos\alpha Re(V_{k,1} U_{k,2}) +
  \sin\alpha Re(V_{k,2} U_{k,1}) \right]
\eqno{(A.12)}
$$
$$
V_{h^0\tilde{\chi}^{+}_{k}\tilde{\chi}^{+}_{k}}^{s} =
   \frac{i g}{\sqrt{2}} \left[ \sin\alpha Re(V_{k,1} U_{k,2}) -
   \cos\alpha Re(V_{k,2} U_{k,1}) \right]
\eqno{(A.13)}
$$
$$
V_{A^0\tilde{\chi}^{+}_{k}\tilde{\chi}^{+}_{k}}^{ps} =
   \frac{g}{\sqrt{2}} \left[ \sin\beta Re(V_{k,1} U_{k,2}) +
   \cos\beta Re(V_{k,2} U_{k,1}) \right]
\eqno{(A.14)}
$$
$$
V_{G^0\tilde{\chi}^{+}_{k}\tilde{\chi}^{+}_{k}}^{ps} =
   \frac{-g}{\sqrt{2}} \left[ \cos\beta Re(V_{k,1} U_{k,2}) -
   \sin\beta Re(V_{k,2} U_{k,1}) \right]
\eqno{(A.15)}
$$

{\large B. Form Factors.}
In this appendix we list all the form factors for the one-loop correction
diagrams by using some abbreviations for following expressions.
$$
\bar{B}_{0}^{1,k}=B_{0}[-p_1-p_2,m_{\tilde{b}_k},m_{\tilde{b}_k}]-\Delta,~~~
\bar{B}_{0}^{2,k}=\bar{B}_{0}^{1,k}(m_{\tilde{b}_k} \rightarrow m_{\tilde{t}_k}),
$$
$$
\bar{B}_{0}^{3,i,j}=B_{0}[p_3-p_1, m_{\tilde{\chi}_i^0},
                             m_{\tilde{t}_j}]-\Delta,~~~
\bar{B}_{1}^{3,i,j}=B_{0}[p_3-p_1, m_{\tilde{\chi}_i^0},
                             m_{\tilde{t}_j}]+\frac{\Delta}{2},
$$
$$
\bar{B}_{0}^{4,i,j}=B_{0}[p_3-p_1, m_{\tilde{\chi}_i^+},
                             m_{\tilde{b}_j}]-\Delta,~~~
\bar{B}_{1}^{4,i,j}=B_{0}[p_3-p_1, m_{\tilde{\chi}_i^+},
                             m_{\tilde{b}_j}]+\frac{\Delta}{2},
$$
$$
C_{0}^{1,i,j},C_{ab}^{1,i,j}=
    C_{0},C_{ab}[-p_1,p_1+p_2,m_{\tilde{\chi}_i^+},
                              m_{\tilde{b}_j},m_{\tilde{b}_j}],
$$
$$
C_{0}^{2,i,j},C_{ab}^{2,i,j}=
    C_{0},C_{ab}[-p_1,p_1+p_2,m_{\tilde{\chi}_i^0},
                              m_{\tilde{t}_j},m_{\tilde{t}_j}],
$$
$$
C_{0}^{3,k},C_{ab}^{3,k}=
    C_{0},C_{ab}[-p_3,p_1+p_2,m_{\tilde{\chi}_k^+},
                              m_{\tilde{\chi}_k^+},m_{\tilde{\chi}_k^+}],
$$
$$
C_{0}^{4,k},C_{ab}^{4,k}=
    C_{0},C_{ab}[p_3,-p_1-p_2,m_{\tilde{b}_k},
                              m_{\tilde{b}_k},m_{\tilde{b}_k}],
$$
$$
C_{0}^{5,k},C_{ab}^{5,k}=
    C_{0},C_{ab}[p_3,-p_1-p_2,m_{\tilde{t}_k},
                              m_{\tilde{t}_k},m_{\tilde{t}_k}],
$$
$$
C_{0}^{6,i,j},C_{ab}^{6,i,j}(k_1, k_2) =
    C_{0},C_{ab} [-k_1, k_1 + k_2, m_{\tilde{b}_j},
      m_{\tilde{\chi}_i^+}, m_{\tilde{\chi}_i^+}],
$$
$$
C_{0}^{7,i,j},C_{ab}^{7,i,j}(k_1, k_2) =
    C_{0},C_{ab} [-k_1, k_1 + k_2, m_{\tilde{\chi}_i^+},
      m_{\tilde{b}_j}, m_{\tilde{b}_j}],
$$
$$
C_{0}^{8,i,j},C_{ab}^{8,i,j}(k_1, k_2) =
    C_{0},C_{ab} [-k_1, k_1 + k_2, m_{\tilde{\chi}_i^0},
      m_{\tilde{t}_j}, m_{\tilde{t}_j}],
$$
$$
D_{0}^{1,i,j},D_{ab}^{1,i,j},D_{abc}^{1,i,j}=
    D_{0},D_{ab},D_{abc}[p_1,-p_3,-p_4,m_{\tilde{b}_j},m_{\tilde{\chi}_i^+},
                                   m_{\tilde{\chi}_i^+},m_{\tilde{\chi}_i^+}]
$$
$$
D_{0}^{2,i,j},D_{ab}^{2,i,j},D_{abc}^{2,i,j}=
    D_{0},D_{ab},D_{abc}[-p_1,p_3,p_4,m_{\tilde{\chi}_i^+},
                             m_{\tilde{b}_j},m_{\tilde{b}_j},m_{\tilde{b}_j}]
$$
$$
D_{0}^{3,i,j},D_{ab}^{3,i,j},D_{abc}^{3,i,j}=
    D_{0},D_{ab},D_{abc}[-p_1,p_3,p_4,m_{\tilde{\chi}_i^0},
                             m_{\tilde{t}_j},m_{\tilde{t}_j},m_{\tilde{t}_j}]
$$
$$
D_{0}^{4,i,j},D_{ab}^{4,i,j},D_{abc}^{4,i,j}=
    D_{0},D_{ab},D_{abc}[-p_3,p_1,-p_4,m_{\tilde{b}_j},m_{\tilde{b}_j},
                                   m_{\tilde{\chi}_i^+},m_{\tilde{\chi}_i^+}]
$$
\begin{eqnarray*}
A_t=\frac{i}{\hat{t}-m_t^2},~~~~
A_u=\frac{i}{\hat{u}-m_t^2},
\end{eqnarray*}
\begin{eqnarray*}
A_h=\frac{i}{\hat{s}-m_{h}^2},~~~~~
A_H=\frac{i}{\hat{s}-m_{H}^2}.
\end{eqnarray*}
\begin{eqnarray*}
A_A=\frac{i}{\hat{s}-m_{A}^2},~~~~~
A_G=\frac{i}{\hat{s}-m_{Z}^2}.
\end{eqnarray*}
$$
F_{1}^{t\tilde{b}_j\tilde{\chi}_i^+}=
  -|V_{t\tilde{b}_j\tilde{\chi}_i^+}^{(1)}|^2-
   |V_{t\tilde{b}_j\tilde{\chi}_i^+}^{(2)}|^2,~~~
F_{2}^{t\tilde{b}_j\tilde{\chi}_i^+}=
   -V_{t\tilde{b}_j\tilde{\chi}_i^+}^{(1)\ast}
    V_{t\tilde{b}_j\tilde{\chi}_i^+}^{(2)}-
    V_{t\tilde{b}_j\tilde{\chi}_i^+}^{(2)\ast}
    V_{t\tilde{b}_j\tilde{\chi}_i^+}^{(1)},
$$
and
$$
G_{1}^{t\tilde{t}_j\tilde{\chi}_i^0}=
  -|V_{t\tilde{t}_j\tilde{\chi}_i^0}^{(1)}|^2-
   |V_{t\tilde{t}_j\tilde{\chi}_i^0}^{(2)}|^2,~~~
G_{2}^{t\tilde{t}_j\tilde{\chi}_i^0}=
   -V_{t\tilde{t}_j\tilde{\chi}_i^0}^{(1)\ast}
    V_{t\tilde{t}_j\tilde{\chi}_i^0}^{(2)}-
    V_{t\tilde{t}_j\tilde{\chi}_i^0}^{(2)\ast}
    V_{t\tilde{t}_j\tilde{\chi}_i^0}^{(1)}.
$$
\par
The one-particle-irreducible(1PI) correction to the vertex $\gamma tt$
stemming from squark, chargino and neutralino can be written in terms of
form factors
\begin{eqnarray*}
\Delta\Gamma_{\gamma tt}^{\mu}(k_1,k_2)&=&
   g_{1}(k_1,k_2) k_{1}^{\mu} \gamma_5 \rlap/{k}_1
   + g_{2}(k_1,k_2) k_{2}^{\mu} \gamma_5 \rlap/{k}_1
   + g_{3}(k_1,k_2) k_{1}^{\mu} \gamma_5 \rlap/{k}_2 \\
&+& g_{4}(k_1,k_2) k_{2}^{\mu} \gamma_5 \rlap/{k}_2
   + g_{5}(k_1,k_2) k_{1}^{\mu} \gamma_5
   + g_{6}(k_1,k_2) k_{2}^{\mu} \gamma_5  \\
&+& g_{7}(k_1,k_2) \gamma_5 \gamma^{\mu} \rlap/{k}_1 \rlap/{k}_2
   + g_{8}(k_1,k_2) \gamma_5 \gamma^{\mu} \rlap/{k}_1
   + g_{9}(k_1,k_2) \gamma_5 \gamma^{\mu} \rlap/{k}_2 \\
&+& g_{10}(k_1,k_2) \gamma_5 \gamma^{\mu}
   + g_{11}(k_1,k_2) k_{1}^{\mu} \rlap/{k}_1
   + g_{12}(k_1,k_2) k_{2}^{\mu} \rlap/{k}_1 \\
&+& g_{13}(k_1,k_2) k_{1}^{\mu} \rlap/{k}_2
   + g_{14}(k_1,k_2) k_{2}^{\mu} \rlap/{k}_2
   + g_{15}(k_1,k_2) k_{1}^{\mu} \\
&+& g_{16}(k_1,k_2) k_{2}^{\mu}
   + g_{17}(k_1,k_2) \gamma^{\mu} \rlap/{k}_1 \rlap/{k}_2
   + g_{18}(k_1,k_2) \gamma^{\mu} \rlap/{k}_1 \\
&+& g_{19}(k_1,k_2) \gamma^{\mu} \rlap/{k}_2
   + g_{20}(k_1,k_2) \gamma^{\mu},
\end{eqnarray*}
where $k_1$ and $k_2$ are the four-momenta of the lightest top quark pair and
along their outgoing directions, respectively. In the equation above, the form
factors of the Lorentz invariant structures including $\gamma_5$ do not
contribute to the cross sections of our subprocess. Therefore we shall list
only the explicit expressions of the form factors $g_{i}~(i=11 \sim 20)$.
The form factors $g_{i}~(i=11 \sim 20)$ are expressed as follows.
\begin{eqnarray*}
g_{11}(k_1, k_2) &=& \frac{i e}{16 \pi^2} \sum_{i=1,2} \sum_{j=1,2}
      F_{1}^{t\tilde{b}_j\tilde{\chi}_i^+} (C^{6,i,j}_{11} - C^{6,i,j}_{12} +
      C^{6,i,j}_{21} + C^{6,i,j}_{22} - 2 C^{6,i,j}_{23}) (k_1, k_2) \\
&-&
      \frac{i e}{32 \pi^2} Q_b \sum_{i=1,2} \sum_{j=1,2}
      F_{1}^{t\tilde{b}_j\tilde{\chi}_i^+} (C^{7,i,j}_{11} - C^{7,i,j}_{12} +
      2 C^{7,i,j}_{21} + 2 C^{7,i,j}_{22} - 4 C^{7,i,j}_{23}) (k_1, k_2) \\
&-&
      \frac{i e}{32 \pi^2} Q_t \sum_{i=1,4} \sum_{j=1,2}
      G_{1}^{t\tilde{t}_j\tilde{\chi}_i^0} (C^{8,i,j}_{11} - C^{8,i,j}_{12} +
      2 C^{8,i,j}_{21} + 2 C^{8,i,j}_{22} - 4 C^{8,i,j}_{23}) (k_1, k_2)
\end{eqnarray*}
\begin{eqnarray*}
g_{12}(k_1, k_2) &=& \frac{i e}{16 \pi^2} \sum_{i=1,2} \sum_{j=1,2}
      F_{1}^{t\tilde{b}_j\tilde{\chi}_i^+} (C^{6,i,j}_{22} - C^{6,i,j}_{23})
      (k_1, k_2) \\
&+&
      \frac{i e}{32 \pi^2} Q_b \sum_{i=1,2} \sum_{j=1,2}
      F_{1}^{t\tilde{b}_j\tilde{\chi}_i^+} (C^{7,i,j}_{11} - C^{7,i,j}_{12} +
      2 C^{7,i,j}_{22} + 2 C^{7,i,j}_{23}) (k_1, k_2) \\
&+&
      \frac{i e}{32 \pi^2} Q_t \sum_{i=1,4} \sum_{j=1,2}
      G_{1}^{t\tilde{t}_j\tilde{\chi}_i^0} (C^{8,i,j}_{11} - C^{8,i,j}_{12} +
      2 C^{8,i,j}_{22} + 2 C^{8,i,j}_{23}) (k_1, k_2)
\end{eqnarray*}
\begin{eqnarray*}
g_{13}(k_1, k_2) &=& \frac{-i e}{16 \pi^2} \sum_{i=1,2} \sum_{j=1,2}
      F_{1}^{t\tilde{b}_j\tilde{\chi}_i^+} (C^{6,i,j}_0 + C^{6,i,j}_{11} -
      C^{6,i,j}_{22} + C^{6,i,j}_{23}) (k_1, k_2) \\
&+&
      \frac{i e}{32 \pi^2} Q_b \sum_{i=1,2} \sum_{j=1,2}
      F_{1}^{t\tilde{b}_j\tilde{\chi}_i^+} (C^{7,i,j}_{12} -
      2 C^{7,i,j}_{22} + 2 C^{7,i,j}_{23}) (k_1, k_2) \\
&+&
      \frac{i e}{32 \pi^2} Q_t \sum_{i=1,4} \sum_{j=1,2}
      G_{1}^{t\tilde{t}_j\tilde{\chi}_i^0} (C^{8,i,j}_{12} -
      2 C^{8,i,j}_{22} + 2 C^{8,i,j}_{23}) (k_1, k_2)
\end{eqnarray*}
\begin{eqnarray*}
g_{14}(k_1, k_2) &=& \frac{i e}{16 \pi^2} \sum_{i=1,2} \sum_{j=1,2}
      F_{1}^{t\tilde{b}_j\tilde{\chi}_i^+} (C^{6,i,j}_{12} + C^{6,i,j}_{22})
      (k_1, k_2) \\
&-&
      \frac{i e}{32 \pi^2} Q_b \sum_{i=1,2} \sum_{j=1,2}
      F_{1}^{t\tilde{b}_j\tilde{\chi}_i^+} (C^{7,i,j}_{12} + 2 C^{7,i,j}_{22})
      (k_1, k_2) \\
&-&
      \frac{i e}{32 \pi^2} Q_t \sum_{i=1,4} \sum_{j=1,2}
      G_{1}^{t\tilde{t}_j\tilde{\chi}_i^0} (C^{8,i,j}_{12} + 2 C^{8,i,j}_{22})
      (k_1, k_2)
\end{eqnarray*}
\begin{eqnarray*}
g_{15}(k_1, k_2) &=& \frac{i e}{16 \pi^2} \sum_{i=1,2} \sum_{j=1,2}
      F_{2}^{t\tilde{b}_j\tilde{\chi}_i^+} m_{\tilde{\chi}_i^+}
      (C^{6,i,j}_0 + C^{6,i,j}_{11} - C^{6,i,j}_{12})
      (k_1, k_2) \\
&+&
      \frac{i e}{32 \pi^2} Q_b \sum_{i=1,2} \sum_{j=1,2}
      F_{2}^{t\tilde{b}_j\tilde{\chi}_i^+} m_{\tilde{\chi}_i^+}
      (C^{7,i,j}_0 + 2 C^{7,i,j}_{11} - 2 C^{7,i,j}_{12})
      (k_1, k_2) \\
&+&
      \frac{i e}{32 \pi^2} Q_t \sum_{i=1,4} \sum_{j=1,2}
      G_{2}^{t\tilde{t}_j\tilde{\chi}_i^0} m_{\tilde{\chi}_i^0}
      (C^{8,i,j}_0 + 2 C^{8,i,j}_{11} - 2 C^{8,i,j}_{12})
      (k_1, k_2)
\end{eqnarray*}
\begin{eqnarray*}
g_{16}(k_1, k_2) &=& \frac{-i e}{16 \pi^2} \sum_{i=1,2} \sum_{j=1,2}
      F_{2}^{t\tilde{b}_j\tilde{\chi}_i^+} m_{\tilde{\chi}_i^+}
      C^{6,i,j}_{12} (k_1, k_2) \\
&-&
      \frac{i e}{32 \pi^2} Q_b \sum_{i=1,2} \sum_{j=1,2}
      F_{2}^{t\tilde{b}_j\tilde{\chi}_i^+} m_{\tilde{\chi}_i^+}
      (C^{7,i,j}_0 + 2 C^{7,i,j}_{12}) (k_1, k_2) \\
&-&
      \frac{i e}{32 \pi^2} Q_t \sum_{i=1,4} \sum_{j=1,2}
      G_{2}^{t\tilde{t}_j\tilde{\chi}_i^0} m_{\tilde{\chi}_i^0}
      (C^{8,i,j}_0 + 2 C^{8,i,j}_{12}) (k_1, k_2)
\end{eqnarray*}
\begin{eqnarray*}
g_{17}(k_1, k_2) = \frac{i e}{32 \pi^2} \sum_{i=1,2} \sum_{j=1,2}
      F_{1}^{t\tilde{b}_j\tilde{\chi}_i^+} (C^{6,i,j}_0 + C^{6,i,j}_{11})
      (k_1, k_2)
\end{eqnarray*}
\begin{eqnarray*}
g_{18}(k_1, k_2) = g_{19}(k_1, k_2) =
      \frac{-i e}{32 \pi^2} \sum_{i=1,2} \sum_{j=1,2}
      F_{2}^{t\tilde{b}_j\tilde{\chi}_i^+} m_{\tilde{\chi}_i^+}
      C^{6,i,j}_0 (k_1, k_2)
\end{eqnarray*}
\begin{eqnarray*}
g_{20}(k_1, k_2) &=& \frac{-i e}{32 \pi^2} \sum_{i=1,2} \sum_{j=1,2}
      F_{1}^{t\tilde{b}_j\tilde{\chi}_i^+} ((\epsilon-2) C^{6,i,j}_{24} +
      (k_1 \cdot k_1) (C^{6,i,j}_{11} - C^{6,i,j}_{12} + C^{6,i,j}_{21} \\
&&  + C^{6,i,j}_{22} - 2 C^{6,i,j}_{23}) + 2 (k_1 \cdot k_2)
      (C^{6,i,j}_{22} - C^{6,i,j}_{23}) + (k_2 \cdot k_2) (C^{6,i,j}_{12} +
      C^{6,i,j}_{22}) \\
&&  - m_{\tilde{\chi}_i^+}^2 C^{6,i,j}_0) (k_1, k_2) \\
&+&   \frac{i e}{16 \pi^2} Q_b \sum_{i=1,2} \sum_{j=1,2}
      F_{1}^{t\tilde{b}_j\tilde{\chi}_i^+} C^{7,i,j}_{24} (k_1, k_2) +
      \frac{i e}{16 \pi^2} Q_t \sum_{i=1,4} \sum_{j=1,2}
      G_{1}^{t\tilde{t}_j\tilde{\chi}_i^0} C^{8,i,j}_{24} (k_1, k_2)
\end{eqnarray*}
\par
Then the form factors in the renormalized amplitude of the t-channel vertex
diagrams in the process $\gamma\gamma \rightarrow t\bar{t}$ can be written as:
\begin{eqnarray*}
f_{i}^{v,\hat{t}}=0~~(i=2,3,6,7,9,10,12,13,16 \sim 22),
\end{eqnarray*}
\begin{eqnarray*}
f_1^{v,\hat{t}}&=& 2 i e Q_t (p_1 \cdot p_3) A_{t} \left\{
  m_t \left[ g_{17}(p_1,p_3-p_1) +g_{17}(p_1-p_3,p_2) \right]
  \right. \\
&-& \left. g_{18}(p_1-p_3,p_2)-g_{19}(p_1,p_3-p_1) \right\},
\end{eqnarray*}
\begin{eqnarray*}
f_4^{v,\hat{t}}=2 i e Q_t (p_1 \cdot p_3) A_{t} \left[
       g_{12}(p_1-p_3,p_2)-g_{11}(p_1-p_3,p_2) \right],
\end{eqnarray*}
\begin{eqnarray*}
f_5^{v,\hat{t}}&=&-2 i e Q_t A_{t} \left\{
      i e \left[ C^{+}+C^{-} \right] +m_t^2 (g_{11}(p_1,p_3-p_1)-
     g_{12}(p_1,p_3-p_1)\right. \\
&-&\left. g_{13}(p_1,p_3-p_1)+g_{14}(p_1,p_3-p_1)-g_{17}(p_1,p_3-p_1)+
     g_{17}(p_1-p_3,p_2)) \right. \\
&+&\left. (p_1 \cdot p_3) (g_{13}(p_1,p_3-p_1)- g_{14}(p_1,p_3-p_1)+
     2 g_{17}(p_1,p_3-p_1)) \right. \\
&+&\left. m_t (g_{15}(p_1,p_3-p_1)-g_{16}(p_1,p_3-p_1)
     +g_{18}(p_1,p_3-p_1)-g_{18}(p_1-p_3,p_2) \right. \\
&-&\left. g_{19}(p_1,p_3-p_1)-
     g_{19}(p_1-p_3,p_2) )+g_{20}(p_1,p_3-p_1)+g_{20}(p_1-p_3,p_2) \right\},
\end{eqnarray*}
\begin{eqnarray*}
f_8^{v,\hat{t}}&=& 2 f_{14}^{v,\hat{t}} =2 i e Q_t A_{t} \left\{
     m_t \left[ g_{11}(p_1-p_3,p_2)-g_{12}(p_1-p_3,p_2)-
     g_{13}(p_1-p_3,p_2) \right.\right.\\
&+&\left. \left. g_{14}(p_1-p_3,p_2)-2 g_{17}(p_1-p_3,p_2) \right]+
     g_{15}(p_1-p_3,p_2)- g_{16}(p_1-p_3,p_2) \right. \\
&+&\left. 2 g_{18}(p_1-p_3,p_2) \right\},
\end{eqnarray*}
\begin{eqnarray*}
f_{11}^{v,\hat{t}}&=&-i e Q_t A_{t} \left\{
     i e (C^{+}+C^{-})+m_t^2 (g_{17}(p_1,p_3-p_1)+
     g_{17}(p_1-p_3,p_2)) \right. \\
&-& \left. m_t (g_{18}(p_1,p_3-p_1)+g_{18}(p_1-p_3,p_2)
     +g_{19}(p_1,p_3-p_1) \right. \\
&+& \left. g_{19}(p_1-p_3,p_2))+g_{20}(p_1,p_3-p_1)+g_{20}(p_1-p_3,p_2)
    \right\},
\end{eqnarray*}
\begin{eqnarray*}
f_{15}^{v,\hat{t}}=
     f_{14}^{v,\hat{t}}(g_{i}(p_1-p_3,p_2) \rightarrow g_{i}(p_1,p_3-p_1)).
\end{eqnarray*}
\par
The form factors from the renormalized amplitude of t-channel box
diagrams(Fig.1(c)) are expressed as:
\begin{eqnarray*}
f_1^{b,\hat{t}} &=&
\frac{-i e^2}{32 \pi^2} \sum_{i=1,2} \sum_{j=1,2} (
    F_{2}^{t\tilde{b}_j\tilde{\chi}_i^+} m_{\tilde{\chi}_i^+} \left[ 2
    p_1 \cdot p_2 (D^{1,i,j}_{13}+D^{1,i,j}_{25}-D^{1,i,j}_{23}) \right.\\
&+& 2 p_1 \cdot p_3 (D^{1,i,j}_{11}+D^{1,i,j}_{12}+D^{1,i,j}_{23}+
    D^{1,i,j}_{24}-D^{1,i,j}_{25}-D^{1,i,j}_{26}-2 D^{1,i,j}_{13})\\
&+& 2 p_2 \cdot p_3 (D^{1,i,j}_{23}-D^{1,i,j}_{13}-D^{1,i,j}_{26})+m_t^2 (2
    D^{1,i,j}_{13}+2 D^{1,i,j}_{25}-2 D^{1,i,j}_{11}-2 D^{1,i,j}_{23}\\
&-& \left. D^{1,i,j}_{0}-D^{1,i,j}_{21})+2 D^{1,i,j}_{27}+m_{\tilde{\chi}_i^+}^
    2 D^{1,i,j}_{0} \right]+F_{1}^{t\tilde{b}_j\tilde{\chi}_i^+} \left\{ 2
    p_1 \cdot p_2 m_t (D^{1,i,j}_{13} \right.\\
&+& D^{1,i,j}_{35}+2 D^{1,i,j}_{25}-D^{1,i,j}_{23}-D^{1,i,j}_{37})+2
    p_1 \cdot p_3 m_t (D^{1,i,j}_{11}+D^{1,i,j}_{12}+D^{1,i,j}_{21}\\
&+& D^{1,i,j}_{23}+D^{1,i,j}_{34}+D^{1,i,j}_{37}+2 D^{1,i,j}_{24}-3
    D^{1,i,j}_{25}-D^{1,i,j}_{26}-D^{1,i,j}_{310}-D^{1,i,j}_{35}-2
    D^{1,i,j}_{13})\\
&+& 2 p_2 \cdot p_3 m_t (D^{1,i,j}_{23}+D^{1,i,j}_{37}-D^{1,i,j}_{13}-
    D^{1,i,j}_{25}-D^{1,i,j}_{26}-D^{1,i,j}_{310})\\
&+& m_t m_{\tilde{\chi}_i^+}^2 (D^{1,i,j}_{0}+D^{1,i,j}_{11})+m_t \left[ 4
    D^{1,i,j}_{27}+(4-\epsilon) D^{1,i,j}_{311} \right] \\
&+& m_t^3 (2 D^{1,i,j}_{13}+2 D^{1,i,j}_{35}+4
    D^{1,i,j}_{25}-3 D^{1,i,j}_{11}-3 D^{1,i,j}_{21}-2 D^{1,i,j}_{23}-2
    D^{1,i,j}_{37}\\
&-& \left. D^{1,i,j}_{0}-D^{1,i,j}_{31}) \right\})-
    \frac{i e^2 Q_b^2}{16 \pi^2} \sum_{i=1,2} \sum_{j=1,2} (
    F_{1}^{t\tilde{b}_j\tilde{\chi}_i^+} m_t D^{2,i,j}_{311}-
    F_{2}^{t\tilde{b}_j\tilde{\chi}_i^+} m_{\tilde{\chi}_i^+} D^{2,i,j}_{27} )\\
&-& \frac{i e^2 Q_t^2}{16 \pi^2} \sum_{i=1,4} \sum_{j=1,2} (
    G_{1}^{t\tilde{t}_j\tilde{\chi}_i^0} m_t D^{3,i,j}_{311}-
    G_{2}^{t\tilde{t}_j\tilde{\chi}_i^0} m_{\tilde{\chi}_i^0} D^{3,i,j}_{27} )\\
&-& \frac{i e^2 Q_b}{16 \pi^2} \sum_{i=1,2} \sum_{j=1,2} \left[
    F_{2}^{t\tilde{b}_j\tilde{\chi}_i^+} m_{\tilde{\chi}_i^+} D^{4,i,j}_{27}+
    F_{1}^{t\tilde{b}_j\tilde{\chi}_i^+} m_t (D^{4,i,j}_{27}+D^{4,i,j}_{312})
     \right]
\end{eqnarray*}

\begin{eqnarray*}
f_2^{b,\hat{t}} &=&
\frac{i e^2}{16 \pi^2} \sum_{i=1,2} \sum_{j=1,2} \left[
    F_{1}^{t\tilde{b}_j\tilde{\chi}_i^+} m_t (D^{1,i,j}_{27}+D^{1,i,j}_{311})+
    F_{2}^{t\tilde{b}_j\tilde{\chi}_i^+} m_{\tilde{\chi}_i^+} D^{1,i,j}_{27}
     \right]\\
&-& \frac{i e^2 Q_b^2}{16 \pi^2} \sum_{i=1,2} \sum_{j=1,2} (
    F_{1}^{t\tilde{b}_j\tilde{\chi}_i^+} m_t D^{2,i,j}_{311}-
    F_{2}^{t\tilde{b}_j\tilde{\chi}_i^+} m_{\tilde{\chi}_i^+} D^{2,i,j}_{27} )\\
&-& \frac{i e^2 Q_t^2}{16 \pi^2} \sum_{i=1,4} \sum_{j=1,2} (
    G_{1}^{t\tilde{t}_j\tilde{\chi}_i^0} m_t D^{3,i,j}_{311}-
    G_{2}^{t\tilde{t}_j\tilde{\chi}_i^0} m_{\tilde{\chi}_i^0} D^{3,i,j}_{27} )\\
&-& \frac{-i e^2 Q_b}{16 \pi^2} \sum_{i=1,2} \sum_{j=1,2} \left[
    F_{2}^{t\tilde{b}_j\tilde{\chi}_i^+} m_{\tilde{\chi}_i^+} D^{4,i,j}_{27}+
    F_{1}^{t\tilde{b}_j\tilde{\chi}_i^+} m_t (D^{4,i,j}_{27}+D^{4,i,j}_{312})
     \right]
\end{eqnarray*}

\begin{eqnarray*}
f_3^{b,\hat{t}} &=&
\frac{-i e^2}{16 \pi^2} \sum_{i=1,2} \sum_{j=1,2}
    F_{1}^{t\tilde{b}_j\tilde{\chi}_i^+} \left[ 2 p_1 \cdot p_2 (D^{1,i,j}_{25}
    +D^{1,i,j}_{35}+D^{1,i,j}_{39}-D^{1,i,j}_{26}-D^{1,i,j}_{310}-
    D^{1,i,j}_{37}) \right.\\
&+& 2 p_1 \cdot p_3 (D^{1,i,j}_{24}+D^{1,i,j}_{26}+D^{1,i,j}_{34}+
    D^{1,i,j}_{37}+D^{1,i,j}_{38}-D^{1,i,j}_{22}-D^{1,i,j}_{25}-D^{1,i,j}_{35}
    \\
&-& D^{1,i,j}_{36}-D^{1,i,j}_{39})+2 p_2 \cdot p_3 (D^{1,i,j}_{26}+
    D^{1,i,j}_{37}+D^{1,i,j}_{38}-D^{1,i,j}_{25}-D^{1,i,j}_{310}-D^{1,i,j}_{39}
    )\\
&+& m_t^2 (D^{1,i,j}_{12}+D^{1,i,j}_{34}+2 D^{1,i,j}_{24}+2 D^{1,i,j}_{25}+2
    D^{1,i,j}_{35}+2 D^{1,i,j}_{39}-D^{1,i,j}_{11}-D^{1,i,j}_{31}\\
&-& 2 D^{1,i,j}_{21}-2 D^{1,i,j}_{26}-2 D^{1,i,j}_{310}-2 D^{1,i,j}_{37})+
    m_{\tilde{\chi}_i^+}^2 (D^{1,i,j}_{11}-D^{1,i,j}_{12})\\
&+& \left. (4-\epsilon) (D^{1,i,j}_{311}-D^{1,i,j}_{312}) \right]-
    \frac{i e^2 Q_b^2}{8 \pi^2} \sum_{i=1,2} \sum_{j=1,2}
    F_{1}^{t\tilde{b}_j\tilde{\chi}_i^+} (D^{2,i,j}_{311}-D^{2,i,j}_{312})\\
&-& \frac{i e^2 Q_t^2}{8 \pi^2} \sum_{i=1,4} \sum_{j=1,2}
    G_{1}^{t\tilde{t}_j\tilde{\chi}_i^0} (D^{3,i,j}_{311}-D^{3,i,j}_{312})-
    \frac{i e^2 Q_b}{8 \pi^2} \sum_{i=1,2} \sum_{j=1,2}
    F_{1}^{t\tilde{b}_j\tilde{\chi}_i^+} (D^{4,i,j}_{312}-D^{4,i,j}_{311})
\end{eqnarray*}

\begin{eqnarray*}
f_4^{b,\hat{t}} &=&
\frac{-i e^2}{16 \pi^2} \sum_{i=1,2} \sum_{j=1,2}
    F_{1}^{t\tilde{b}_j\tilde{\chi}_i^+} \left[ 2 p_1 \cdot p_2 (D^{1,i,j}_{23}
    +D^{1,i,j}_{37}+D^{1,i,j}_{39}-D^{1,i,j}_{26}-D^{1,i,j}_{310}-
    D^{1,i,j}_{33}) \right.\\
&+& 2 p_1 \cdot p_3 (2 D^{1,i,j}_{26}+2 D^{1,i,j}_{310}+D^{1,i,j}_{25}+
    D^{1,i,j}_{33}+D^{1,i,j}_{38}-2 D^{1,i,j}_{23}-2 D^{1,i,j}_{39}-
    D^{1,i,j}_{22}\\
&-& D^{1,i,j}_{36}-D^{1,i,j}_{37})+2 p_2 \cdot p_3 (D^{1,i,j}_{26}+
    D^{1,i,j}_{33}+D^{1,i,j}_{38}-2 D^{1,i,j}_{39}-D^{1,i,j}_{23})\\
&+& m_t^2 (D^{1,i,j}_{12}+D^{1,i,j}_{34}+2 D^{1,i,j}_{23}+2 D^{1,i,j}_{24}+2
    D^{1,i,j}_{37}+2 D^{1,i,j}_{39}-D^{1,i,j}_{13}-D^{1,i,j}_{35}\\
&-& 2 D^{1,i,j}_{25}-2 D^{1,i,j}_{26}-2 D^{1,i,j}_{310}-2 D^{1,i,j}_{33})+
    m_{\tilde{\chi}_i^+}^2 (D^{1,i,j}_{13}-D^{1,i,j}_{12})\\
&+& \left. (6-\epsilon) D^{1,i,j}_{313}-2 D^{1,i,j}_{27}-(4-\epsilon)
    D^{1,i,j}_{312} \right]+
    \frac{i e^2 Q_b^2}{8 \pi^2} \sum_{i=1,2} \sum_{j=1,2}
    F_{1}^{t\tilde{b}_j\tilde{\chi}_i^+} (D^{2,i,j}_{27}+D^{2,i,j}_{312})\\
&+& \frac{i e^2 Q_t^2}{8 \pi^2} \sum_{i=1,4} \sum_{j=1,2}
    G_{1}^{t\tilde{t}_j\tilde{\chi}_i^0} (D^{3,i,j}_{27}+D^{3,i,j}_{312})+
    \frac{i e^2 Q_b}{8 \pi^2} \sum_{i=1,2} \sum_{j=1,2}
    F_{1}^{t\tilde{b}_j\tilde{\chi}_i^+} D^{4,i,j}_{311}
\end{eqnarray*}

\begin{eqnarray*}
f_5^{b,\hat{t}} &=&
\frac{-i e^2}{16 \pi^2} \sum_{i=1,2} \sum_{j=1,2} \left\{ 2
    F_{2}^{t\tilde{b}_j\tilde{\chi}_i^+} m_t m_{\tilde{\chi}_i^+} (
    D^{1,i,j}_{0}+D^{1,i,j}_{11}-D^{1,i,j}_{13}) \right.\\
&+& F_{1}^{t\tilde{b}_j\tilde{\chi}_i^+} \left[ 2 p_2 \cdot p_3 (D^{1,i,j}_{25}
    -D^{1,i,j}_{26})+m_t^2 (D^{1,i,j}_{0}+D^{1,i,j}_{21}+2 D^{1,i,j}_{11}-2
    D^{1,i,j}_{13} \right.\\
&-& \left. \left. 2 D^{1,i,j}_{25})+m_{\tilde{\chi}_i^+}^2 D^{1,i,j}_{0}+2 (
    D^{1,i,j}_{313}-D^{1,i,j}_{311}) \right] \right\}-
    \frac{i e^2 Q_b^2}{8 \pi^2} \sum_{i=1,2} \sum_{j=1,2}
    F_{1}^{t\tilde{b}_j\tilde{\chi}_i^+} (D^{2,i,j}_{27}+D^{2,i,j}_{311}\\
&-& D^{2,i,j}_{313})-\frac{i e^2 Q_t^2}{8 \pi^2} \sum_{i=1,4} \sum_{j=1,2}
    G_{1}^{t\tilde{t}_j\tilde{\chi}_i^0} (D^{3,i,j}_{27}+D^{3,i,j}_{311}-
    D^{3,i,j}_{313})-\frac{i e^2 Q_b}{16 \pi^2} \sum_{i=1,2} \sum_{j=1,2}
    \left\{  \right. \\
&&  2 F_{2}^{t\tilde{b}_j\tilde{\chi}_i^+} m_t m_{\tilde{\chi}_i^+}
    (D^{4,i,j}_{13}-D^{4,i,j}_{12})+
    F_{1}^{t\tilde{b}_j\tilde{\chi}_i^+} \left[ 2 p_1 \cdot p_2 (2
    D^{4,i,j}_{39}-D^{4,i,j}_{33}-D^{4,i,j}_{38}) \right.\\
&+& 2 p_1 \cdot p_3 (2 D^{4,i,j}_{26}+2 D^{4,i,j}_{310}+D^{4,i,j}_{33}+
    D^{4,i,j}_{38}-2 D^{4,i,j}_{39}-D^{4,i,j}_{22}-D^{4,i,j}_{23}-
    D^{4,i,j}_{36}\\
&-& D^{4,i,j}_{37})+2 p_2 \cdot p_3 (D^{4,i,j}_{310}+D^{4,i,j}_{33}-
    D^{4,i,j}_{37}-D^{4,i,j}_{39})+m_t^2 (D^{4,i,j}_{13}+D^{4,i,j}_{32}\\
&+& 4 D^{4,i,j}_{39}-3 D^{4,i,j}_{38}-D^{4,i,j}_{12}-2 D^{4,i,j}_{33})+
    m_{\tilde{\chi}_i^+}^2 (D^{4,i,j}_{13}-D^{4,i,j}_{12})\\
&+& \left. \left. (4-\epsilon) (D^{4,i,j}_{313}-D^{4,i,j}_{312}) \right]
     \right\}
\end{eqnarray*}

\begin{eqnarray*}
f_6^{b,\hat{t}} &=&
\frac{-i e^2}{16 \pi^2} \sum_{i=1,2} \sum_{j=1,2} \left\{ -2
    F_{2}^{t\tilde{b}_j\tilde{\chi}_i^+} m_t m_{\tilde{\chi}_i^+}
    D^{1,i,j}_{13}+F_{1}^{t\tilde{b}_j\tilde{\chi}_i^+} \left[ 2 p_1 \cdot p_2
    (D^{1,i,j}_{33}-D^{1,i,j}_{37}) \right. \right.\\
&+& 2 p_1 \cdot p_3 (D^{1,i,j}_{23}+D^{1,i,j}_{37}+D^{1,i,j}_{39}-
    D^{1,i,j}_{25}-D^{1,i,j}_{310}-D^{1,i,j}_{33})\\
&+& 2 p_2 \cdot p_3 (D^{1,i,j}_{39}-D^{1,i,j}_{33})+m_t^2 (D^{1,i,j}_{35}+2
    D^{1,i,j}_{33}-D^{1,i,j}_{13}-2 D^{1,i,j}_{37})\\
&-& \left. \left. m_{\tilde{\chi}_i^+}^2 D^{1,i,j}_{13}-(4-\epsilon)
    D^{1,i,j}_{313} \right] \right\}+
    \frac{i e^2 Q_b^2}{8 \pi^2} \sum_{i=1,2} \sum_{j=1,2}
    F_{1}^{t\tilde{b}_j\tilde{\chi}_i^+} D^{2,i,j}_{313}+
    \frac{i e^2 Q_t^2}{8 \pi^2} \sum_{i=1,4} \sum_{j=1,2} \\
&&  G_{1}^{t\tilde{t}_j\tilde{\chi}_i^0}
    D^{3,i,j}_{313}-\frac{i e^2 Q_b}{16 \pi^2} \sum_{i=1,2} \sum_{j=1,2}
    \left\{ 2 F_{2}^{t\tilde{b}_j\tilde{\chi}_i^+} m_t m_{\tilde{\chi}_i^+}
    D^{4,i,j}_{13}+F_{1}^{t\tilde{b}_j\tilde{\chi}_i^+} \left[ 2 p_1 \cdot p_2
    (D^{4,i,j}_{39} \right. \right.\\
&-& D^{4,i,j}_{33})+2 p_1 \cdot p_3 (D^{4,i,j}_{26}+D^{4,i,j}_{310}+
    D^{4,i,j}_{33}-D^{4,i,j}_{23}-D^{4,i,j}_{37}-D^{4,i,j}_{39})\\
&+& 2 p_2 \cdot p_3 (D^{4,i,j}_{33}-D^{4,i,j}_{37})+m_t^2 (D^{4,i,j}_{13}+2
    D^{4,i,j}_{39}-D^{4,i,j}_{38}-2 D^{4,i,j}_{33})\\
&+& \left. \left. m_{\tilde{\chi}_i^+}^2 D^{4,i,j}_{13}+(4-\epsilon)
    D^{4,i,j}_{313} \right] \right\}
\end{eqnarray*}

\begin{eqnarray*}
f_7^{b,\hat{t}} &=&
\frac{-i e^2}{8 \pi^2} \sum_{i=1,2} \sum_{j=1,2} \left[
    F_{2}^{t\tilde{b}_j\tilde{\chi}_i^+} m_{\tilde{\chi}_i^+} (D^{1,i,j}_{11}+
    D^{1,i,j}_{21}+D^{1,i,j}_{26}-D^{1,i,j}_{12}-D^{1,i,j}_{24}-D^{1,i,j}_{25})
     \right.\\
&+& F_{1}^{t\tilde{b}_j\tilde{\chi}_i^+} m_t (2 D^{1,i,j}_{21}+D^{1,i,j}_{11}+
    D^{1,i,j}_{26}+D^{1,i,j}_{310}+D^{1,i,j}_{31}-2 D^{1,i,j}_{24}-
    D^{1,i,j}_{12}\\
&-& \left. D^{1,i,j}_{25}-D^{1,i,j}_{34}-D^{1,i,j}_{35}) \right]-
    \frac{i e^2 Q_b^2}{8 \pi^2} \sum_{i=1,2} \sum_{j=1,2} \left[
    F_{2}^{t\tilde{b}_j\tilde{\chi}_i^+} m_{\tilde{\chi}_i^+} (D^{2,i,j}_{11}+
    D^{2,i,j}_{21}+D^{2,i,j}_{26} \right.\\
&-& D^{2,i,j}_{12}-D^{2,i,j}_{24}-D^{2,i,j}_{25})+
    F_{1}^{t\tilde{b}_j\tilde{\chi}_i^+} m_t (D^{2,i,j}_{24}+D^{2,i,j}_{34}+
    D^{2,i,j}_{35}-D^{2,i,j}_{21}\\
&-& \left. D^{2,i,j}_{310}-D^{2,i,j}_{31}) \right]-
    \frac{i e^2 Q_t^2}{8 \pi^2} \sum_{i=1,4} \sum_{j=1,2} \left[
    G_{2}^{t\tilde{t}_j\tilde{\chi}_i^0} m_{\tilde{\chi}_i^0} (D^{3,i,j}_{11}+
    D^{3,i,j}_{21}+D^{3,i,j}_{26}-D^{3,i,j}_{12} \right.\\
&-& D^{3,i,j}_{24}-D^{3,i,j}_{25})+G_{1}^{t\tilde{t}_j\tilde{\chi}_i^0} m_t (
    D^{3,i,j}_{24}+D^{3,i,j}_{34}+D^{3,i,j}_{35}-D^{3,i,j}_{21}-D^{3,i,j}_{310}
    \\
&-& \left. D^{3,i,j}_{31}) \right]-
    \frac{i e^2 Q_b}{8 \pi^2} \sum_{i=1,2} \sum_{j=1,2} \left[
    F_{2}^{t\tilde{b}_j\tilde{\chi}_i^+} m_{\tilde{\chi}_i^+} (D^{4,i,j}_{24}+
    D^{4,i,j}_{26}-D^{4,i,j}_{22}-D^{4,i,j}_{25}) \right.\\
&+& F_{1}^{t\tilde{b}_j\tilde{\chi}_i^+} m_t (D^{4,i,j}_{24}+D^{4,i,j}_{26}+
    D^{4,i,j}_{36}+D^{4,i,j}_{38}-D^{4,i,j}_{22}-D^{4,i,j}_{25}-D^{4,i,j}_{310}
    \\
&-& \left. D^{4,i,j}_{32}) \right]
\end{eqnarray*}

\begin{eqnarray*}
f_8^{b,\hat{t}} &=&
\frac{-i e^2}{8 \pi^2} \sum_{i=1,2} \sum_{j=1,2} \left[
    F_{2}^{t\tilde{b}_j\tilde{\chi}_i^+} m_{\tilde{\chi}_i^+} (-D^{1,i,j}_{12}+
    D^{1,i,j}_{26}-D^{1,i,j}_{24})+F_{1}^{t\tilde{b}_j\tilde{\chi}_i^+} m_t (
    D^{1,i,j}_{26} \right.\\
&+& \left. D^{1,i,j}_{310}-2 D^{1,i,j}_{24}-D^{1,i,j}_{12}-D^{1,i,j}_{34})
     \right]-\frac{i e^2 Q_b^2}{8 \pi^2} \sum_{i=1,2} \sum_{j=1,2} \left[
    F_{2}^{t\tilde{b}_j\tilde{\chi}_i^+} m_{\tilde{\chi}_i^+} (D^{2,i,j}_{13}+
    D^{2,i,j}_{26} \right.\\
&-& D^{2,i,j}_{0}-D^{2,i,j}_{11}-D^{2,i,j}_{12}-D^{2,i,j}_{24})+
    F_{1}^{t\tilde{b}_j\tilde{\chi}_i^+} m_t (D^{2,i,j}_{11}+D^{2,i,j}_{21}+
    D^{2,i,j}_{24}\\
&+& \left. D^{2,i,j}_{34}-D^{2,i,j}_{25}-D^{2,i,j}_{310}) \right]-
    \frac{i e^2 Q_t^2}{8 \pi^2} \sum_{i=1,4} \sum_{j=1,2} \left[
    G_{2}^{t\tilde{t}_j\tilde{\chi}_i^0} m_{\tilde{\chi}_i^0} (D^{3,i,j}_{13}+
    D^{3,i,j}_{26}-D^{3,i,j}_{0} \right.\\
&-& D^{3,i,j}_{11}-D^{3,i,j}_{12}-D^{3,i,j}_{24})+
    G_{1}^{t\tilde{t}_j\tilde{\chi}_i^0} m_t (D^{3,i,j}_{11}+D^{3,i,j}_{21}+
    D^{3,i,j}_{24}+D^{3,i,j}_{34}\\
&-& \left. D^{3,i,j}_{25}-D^{3,i,j}_{310}) \right]-
    \frac{i e^2 Q_b}{8 \pi^2} \sum_{i=1,2} \sum_{j=1,2} \left[
    F_{2}^{t\tilde{b}_j\tilde{\chi}_i^+} m_{\tilde{\chi}_i^+} (D^{4,i,j}_{24}-
    D^{4,i,j}_{25}) \right.\\
&+& \left. F_{1}^{t\tilde{b}_j\tilde{\chi}_i^+} m_t (D^{4,i,j}_{24}+
    D^{4,i,j}_{36}-D^{4,i,j}_{25}-D^{4,i,j}_{310}) \right]
\end{eqnarray*}

\begin{eqnarray*}
f_9^{b,\hat{t}} &=&
\frac{-i e^2}{8 \pi^2} \sum_{i=1,2} \sum_{j=1,2} \left[
    F_{2}^{t\tilde{b}_j\tilde{\chi}_i^+} m_{\tilde{\chi}_i^+} (D^{1,i,j}_{26}-
    D^{1,i,j}_{25})+F_{1}^{t\tilde{b}_j\tilde{\chi}_i^+} m_t (D^{1,i,j}_{26}+
    D^{1,i,j}_{310} \right.\\
&-& \left. D^{1,i,j}_{25}-D^{1,i,j}_{35}) \right]-
    \frac{i e^2 Q_b^2}{8 \pi^2} \sum_{i=1,2} \sum_{j=1,2} \left[
    F_{2}^{t\tilde{b}_j\tilde{\chi}_i^+} m_{\tilde{\chi}_i^+} (D^{2,i,j}_{26}-
    D^{2,i,j}_{25}) \right.\\
&+& \left. F_{1}^{t\tilde{b}_j\tilde{\chi}_i^+} m_t (D^{2,i,j}_{35}-
    D^{2,i,j}_{310}) \right]-
    \frac{i e^2 Q_t^2}{8 \pi^2} \sum_{i=1,4} \sum_{j=1,2} \left[
    G_{2}^{t\tilde{t}_j\tilde{\chi}_i^0} m_{\tilde{\chi}_i^0} (D^{3,i,j}_{26}-
    D^{3,i,j}_{25}) \right.\\
&+& \left. G_{1}^{t\tilde{t}_j\tilde{\chi}_i^0} m_t (D^{3,i,j}_{35}-
    D^{3,i,j}_{310}) \right]-
    \frac{i e^2 Q_b}{8 \pi^2} \sum_{i=1,2} \sum_{j=1,2} \left[
    F_{2}^{t\tilde{b}_j\tilde{\chi}_i^+} m_{\tilde{\chi}_i^+} (D^{4,i,j}_{26}-
    D^{4,i,j}_{25}) \right.\\
&+& \left. F_{1}^{t\tilde{b}_j\tilde{\chi}_i^+} m_t (D^{4,i,j}_{26}+
    D^{4,i,j}_{38}-D^{4,i,j}_{25}-D^{4,i,j}_{310}) \right]
\end{eqnarray*}

\begin{eqnarray*}
f_{10}^{b,\hat{t}} &=&
\frac{-i e^2}{8 \pi^2} \sum_{i=1,2} \sum_{j=1,2} \left[
    F_{2}^{t\tilde{b}_j\tilde{\chi}_i^+} m_{\tilde{\chi}_i^+} D^{1,i,j}_{26}+
    F_{1}^{t\tilde{b}_j\tilde{\chi}_i^+} m_t (D^{1,i,j}_{26}+D^{1,i,j}_{310})
     \right]\\
&-& \frac{i e^2 Q_b^2}{8 \pi^2} \sum_{i=1,2} \sum_{j=1,2} \left[
    F_{2}^{t\tilde{b}_j\tilde{\chi}_i^+} m_{\tilde{\chi}_i^+} (D^{2,i,j}_{13}+
    D^{2,i,j}_{26})-F_{1}^{t\tilde{b}_j\tilde{\chi}_i^+} m_t (D^{2,i,j}_{25}+
    D^{2,i,j}_{310}) \right]\\
&-& \frac{i e^2 Q_t^2}{8 \pi^2} \sum_{i=1,4} \sum_{j=1,2} \left[
    G_{2}^{t\tilde{t}_j\tilde{\chi}_i^0} m_{\tilde{\chi}_i^0} (D^{3,i,j}_{13}+
    D^{3,i,j}_{26})-G_{1}^{t\tilde{t}_j\tilde{\chi}_i^0} m_t (D^{3,i,j}_{25}+
    D^{3,i,j}_{310}) \right]\\
&+& \frac{i e^2 Q_b}{8 \pi^2} \sum_{i=1,2} \sum_{j=1,2} \left[
    F_{2}^{t\tilde{b}_j\tilde{\chi}_i^+} m_{\tilde{\chi}_i^+} D^{4,i,j}_{25}+
    F_{1}^{t\tilde{b}_j\tilde{\chi}_i^+} m_t (D^{4,i,j}_{25}+D^{4,i,j}_{310})
     \right]
\end{eqnarray*}

\begin{eqnarray*}
f_{11}^{b,\hat{t}} &=&
\frac{-i e^2}{32 \pi^2} \sum_{i=1,2} \sum_{j=1,2} \left\{ 2
    F_{2}^{t\tilde{b}_j\tilde{\chi}_i^+} m_t m_{\tilde{\chi}_i^+} D^{1,i,j}_{0}
    +F_{1}^{t\tilde{b}_j\tilde{\chi}_i^+} \left[ 2 p_1 \cdot p_2 (
    D^{1,i,j}_{25}+D^{1,i,j}_{37} \right. \right.\\
&+& D^{1,i,j}_{39}-D^{1,i,j}_{26}-D^{1,i,j}_{310}-D^{1,i,j}_{33})+2
    p_1 \cdot p_3 (D^{1,i,j}_{33}+D^{1,i,j}_{38}+2 D^{1,i,j}_{26}+2
    D^{1,i,j}_{310}\\
&-& D^{1,i,j}_{22}-D^{1,i,j}_{23}-D^{1,i,j}_{36}-D^{1,i,j}_{37}-2
    D^{1,i,j}_{39})+2 p_2 \cdot p_3 (D^{1,i,j}_{33}+D^{1,i,j}_{38}-2
    D^{1,i,j}_{39})\\
&+& m_t^2 (2 D^{1,i,j}_{24}+2 D^{1,i,j}_{37}+2 D^{1,i,j}_{39}+D^{1,i,j}_{0}+
    D^{1,i,j}_{12}+D^{1,i,j}_{34}-2 D^{1,i,j}_{26}-2 D^{1,i,j}_{310}\\
&-& 2 D^{1,i,j}_{33}-D^{1,i,j}_{13}-D^{1,i,j}_{21}-D^{1,i,j}_{35})+
    m_{\tilde{\chi}_i^+}^2 (D^{1,i,j}_{0}+D^{1,i,j}_{13}-D^{1,i,j}_{12})\\
&+& \left. \left. (4-\epsilon) (D^{1,i,j}_{313}-D^{1,i,j}_{312}) \right]
     \right\}+\frac{-i e^2 Q_b^2}{16 \pi^2} \sum_{i=1,2} \sum_{j=1,2}
    F_{1}^{t\tilde{b}_j\tilde{\chi}_i^+} (D^{2,i,j}_{313}-D^{2,i,j}_{312})\\
&+& \frac{-i e^2 Q_t^2}{16 \pi^2} \sum_{i=1,4} \sum_{j=1,2}
    G_{1}^{t\tilde{t}_j\tilde{\chi}_i^0} (D^{3,i,j}_{313}-D^{3,i,j}_{312})+
    \frac{-i e^2 Q_b}{16 \pi^2} \sum_{i=1,2} \sum_{j=1,2}
    F_{1}^{t\tilde{b}_j\tilde{\chi}_i^+} (D^{4,i,j}_{313}-D^{4,i,j}_{311})
\end{eqnarray*}

\begin{eqnarray*}
f_{12}^{b,\hat{t}} &=&
\frac{-i e^2}{16 \pi^2} \sum_{i=1,2} \sum_{j=1,2}
    F_{1}^{t\tilde{b}_j\tilde{\chi}_i^+} (D^{1,i,j}_{27}+D^{1,i,j}_{312}-
    D^{1,i,j}_{313})+\frac{-i e^2 Q_b^2}{16 \pi^2} \sum_{i=1,2} \sum_{j=1,2}
    F_{1}^{t\tilde{b}_j\tilde{\chi}_i^+} \cdot \\
&&  (D^{2,i,j}_{313}-D^{2,i,j}_{312}) + \frac{-i e^2 Q_t^2}{16 \pi^2}
    \sum_{i=1,4} \sum_{j=1,2}
    G_{1}^{t\tilde{t}_j\tilde{\chi}_i^0} (D^{3,i,j}_{313}-D^{3,i,j}_{312}) \\
&+& \frac{-i e^2 Q_b}{16 \pi^2} \sum_{i=1,2} \sum_{j=1,2}
    F_{1}^{t\tilde{b}_j\tilde{\chi}_i^+} (D^{4,i,j}_{313}-D^{4,i,j}_{27}-
    D^{4,i,j}_{311})
\end{eqnarray*}

\begin{eqnarray*}
f_{13}^{b,\hat{t}} &=&
\frac{-i e^2}{16 \pi^2} \sum_{i=1,2} \sum_{j=1,2} \left[
    F_{2}^{t\tilde{b}_j\tilde{\chi}_i^+} m_{\tilde{\chi}_i^+} (D^{1,i,j}_{11}-
    D^{1,i,j}_{12})+F_{1}^{t\tilde{b}_j\tilde{\chi}_i^+} m_t (D^{1,i,j}_{11}+
    D^{1,i,j}_{21} \right.\\
&-& \left. D^{1,i,j}_{12}-D^{1,i,j}_{24}) \right]
\end{eqnarray*}

\begin{eqnarray*}
f_{14}^{b,\hat{t}} &=&
\frac{i e^2}{16 \pi^2} \sum_{i=1,2} \sum_{j=1,2} \left[
    F_{2}^{t\tilde{b}_j\tilde{\chi}_i^+} m_{\tilde{\chi}_i^+} D^{1,i,j}_{12}+
    F_{1}^{t\tilde{b}_j\tilde{\chi}_i^+} m_t (D^{1,i,j}_{12}+D^{1,i,j}_{24})
     \right]
\end{eqnarray*}

\begin{eqnarray*}
f_{15}^{b,\hat{t}} &=&
\frac{-i e^2}{16 \pi^2} \sum_{i=1,2} \sum_{j=1,2} \left[
    F_{2}^{t\tilde{b}_j\tilde{\chi}_i^+} m_{\tilde{\chi}_i^+} (D^{1,i,j}_{13}-
    D^{1,i,j}_{11})+F_{1}^{t\tilde{b}_j\tilde{\chi}_i^+} m_t (D^{1,i,j}_{13}+
    D^{1,i,j}_{25} \right.\\
&-& \left. D^{1,i,j}_{11}-D^{1,i,j}_{21}) \right]-
    \frac{i e^2 Q_b}{16 \pi^2} \sum_{i=1,2} \sum_{j=1,2} \left[
    F_{2}^{t\tilde{b}_j\tilde{\chi}_i^+} m_{\tilde{\chi}_i^+} (D^{4,i,j}_{12}-
    D^{4,i,j}_{13}) \right.\\
&+& \left. F_{1}^{t\tilde{b}_j\tilde{\chi}_i^+} m_t (D^{4,i,j}_{12}+
    D^{4,i,j}_{22}-D^{4,i,j}_{13}-D^{4,i,j}_{26}) \right]
\end{eqnarray*}

\begin{eqnarray*}
f_{16}^{b,\hat{t}} &=&
\frac{-i e^2}{16 \pi^2} \sum_{i=1,2} \sum_{j=1,2} \left[
    F_{2}^{t\tilde{b}_j\tilde{\chi}_i^+} m_{\tilde{\chi}_i^+} D^{1,i,j}_{13}+
    F_{1}^{t\tilde{b}_j\tilde{\chi}_i^+} m_t (D^{1,i,j}_{13}+D^{1,i,j}_{25})
     \right]\\
&+& \frac{i e^2 Q_b}{16 \pi^2} \sum_{i=1,2} \sum_{j=1,2} \left[
    F_{2}^{t\tilde{b}_j\tilde{\chi}_i^+} m_{\tilde{\chi}_i^+} D^{4,i,j}_{13}+
    F_{1}^{t\tilde{b}_j\tilde{\chi}_i^+} m_t (D^{4,i,j}_{13}+D^{4,i,j}_{26})
     \right]
\end{eqnarray*}

\begin{eqnarray*}
f_{17}^{b,\hat{t}} &=&
\frac{-i e^2}{8 \pi^2} \sum_{i=1,2} \sum_{j=1,2}
    F_{1}^{t\tilde{b}_j\tilde{\chi}_i^+} (D^{1,i,j}_{22}+D^{1,i,j}_{25}+
    D^{1,i,j}_{35}+D^{1,i,j}_{36}+D^{1,i,j}_{39}-D^{1,i,j}_{24}-D^{1,i,j}_{26}
    \\
&-& D^{1,i,j}_{34}-D^{1,i,j}_{37}-D^{1,i,j}_{38})-
    \frac{i e^2 Q_b^2}{8 \pi^2} \sum_{i=1,2} \sum_{j=1,2}
    F_{1}^{t\tilde{b}_j\tilde{\chi}_i^+} (D^{2,i,j}_{24}+D^{2,i,j}_{26}+
    D^{2,i,j}_{34}+D^{2,i,j}_{37}\\
&+& D^{2,i,j}_{38}-D^{2,i,j}_{22}-D^{2,i,j}_{25}-D^{2,i,j}_{35}-D^{2,i,j}_{36}-
    D^{2,i,j}_{39})-\frac{i e^2 Q_t^2}{8 \pi^2} \sum_{i=1,4} \sum_{j=1,2}
    G_{1}^{t\tilde{t}_j\tilde{\chi}_i^0} (D^{3,i,j}_{24}\\
&+& D^{3,i,j}_{26}+D^{3,i,j}_{34}+D^{3,i,j}_{37}+D^{3,i,j}_{38}-D^{3,i,j}_{22}-
    D^{3,i,j}_{25}-D^{3,i,j}_{35}-D^{3,i,j}_{36}-D^{3,i,j}_{39})\\
&-& \frac{i e^2 Q_b}{8 \pi^2} \sum_{i=1,2} \sum_{j=1,2}
    F_{1}^{t\tilde{b}_j\tilde{\chi}_i^+} (D^{4,i,j}_{22}+D^{4,i,j}_{25}+
    D^{4,i,j}_{35}+D^{4,i,j}_{36}+D^{4,i,j}_{39}-D^{4,i,j}_{24}-D^{4,i,j}_{26}
    \\
&-& D^{4,i,j}_{34}-D^{4,i,j}_{37}-D^{4,i,j}_{38})
\end{eqnarray*}

\begin{eqnarray*}
f_{18}^{b,\hat{t}} &=&
\frac{-i e^2}{8 \pi^2} \sum_{i=1,2} \sum_{j=1,2}
    F_{1}^{t\tilde{b}_j\tilde{\chi}_i^+} (D^{1,i,j}_{22}+D^{1,i,j}_{23}+
    D^{1,i,j}_{36}+D^{1,i,j}_{39}-D^{1,i,j}_{25}-D^{1,i,j}_{26}-D^{1,i,j}_{310}
    \\
&-& D^{1,i,j}_{38})-\frac{i e^2 Q_b^2}{8 \pi^2} \sum_{i=1,2} \sum_{j=1,2}
    F_{1}^{t\tilde{b}_j\tilde{\chi}_i^+} (2 D^{2,i,j}_{26}+D^{2,i,j}_{13}+
    D^{2,i,j}_{25}+D^{2,i,j}_{310}+D^{2,i,j}_{38}-D^{2,i,j}_{12}\\
&-& D^{2,i,j}_{22}-D^{2,i,j}_{23}-D^{2,i,j}_{24}-D^{2,i,j}_{36}-D^{2,i,j}_{39})
    -\frac{i e^2 Q_t^2}{8 \pi^2} \sum_{i=1,4} \sum_{j=1,2}
    G_{1}^{t\tilde{t}_j\tilde{\chi}_i^0} (2 D^{3,i,j}_{26}+D^{3,i,j}_{13}\\
&+& D^{3,i,j}_{25}+D^{3,i,j}_{310}+D^{3,i,j}_{38}-D^{3,i,j}_{12}-D^{3,i,j}_{22}
    -D^{3,i,j}_{23}-D^{3,i,j}_{24}-D^{3,i,j}_{36}-D^{3,i,j}_{39})\\
&-& \frac{i e^2 Q_b}{8 \pi^2} \sum_{i=1,2} \sum_{j=1,2}
    F_{1}^{t\tilde{b}_j\tilde{\chi}_i^+} (D^{4,i,j}_{25}+D^{4,i,j}_{26}+
    D^{4,i,j}_{310}+D^{4,i,j}_{35}-D^{4,i,j}_{23}-D^{4,i,j}_{24}-D^{4,i,j}_{34}
    \\
&-& D^{4,i,j}_{37})
\end{eqnarray*}

\begin{eqnarray*}
f_{19}^{b,\hat{t}} &=&
\frac{-i e^2}{8 \pi^2} \sum_{i=1,2} \sum_{j=1,2}
    F_{1}^{t\tilde{b}_j\tilde{\chi}_i^+} (D^{1,i,j}_{25}+D^{1,i,j}_{310}+
    D^{1,i,j}_{39}-D^{1,i,j}_{26}-D^{1,i,j}_{37}-D^{1,i,j}_{38})\\
&-& \frac{i e^2 Q_b^2}{8 \pi^2} \sum_{i=1,2} \sum_{j=1,2}
    F_{1}^{t\tilde{b}_j\tilde{\chi}_i^+} (D^{2,i,j}_{37}+D^{2,i,j}_{38}-
    D^{2,i,j}_{39}-D^{2,i,j}_{310})-
    \frac{i e^2 Q_t^2}{8 \pi^2} \sum_{i=1,4} \sum_{j=1,2}
    G_{1}^{t\tilde{t}_j\tilde{\chi}_i^0} \cdot \\
&&  (D^{3,i,j}_{37} + D^{3,i,j}_{38}-D^{3,i,j}_{39}-D^{3,i,j}_{310})-
    \frac{i e^2 Q_b}{8 \pi^2} \sum_{i=1,2} \sum_{j=1,2}
    F_{1}^{t\tilde{b}_j\tilde{\chi}_i^+} (D^{4,i,j}_{25}+D^{4,i,j}_{35}+
    D^{4,i,j}_{39} \\
&-& D^{4,i,j}_{26} - D^{4,i,j}_{310}-D^{4,i,j}_{37})
\end{eqnarray*}

\begin{eqnarray*}
f_{20}^{b,\hat{t}} &=&
\frac{-i e^2}{8 \pi^2} \sum_{i=1,2} \sum_{j=1,2}
    F_{1}^{t\tilde{b}_j\tilde{\chi}_i^+} (D^{1,i,j}_{23}+D^{1,i,j}_{39}-
    D^{1,i,j}_{26}-D^{1,i,j}_{38})-
    \frac{i e^2 Q_b^2}{8 \pi^2} \sum_{i=1,2} \sum_{j=1,2}
    F_{1}^{t\tilde{b}_j\tilde{\chi}_i^+} \cdot \\
&&  (D^{2,i,j}_{26} + D^{2,i,j}_{38}-D^{2,i,j}_{23}-D^{2,i,j}_{39})-
    \frac{i e^2 Q_t^2}{8 \pi^2} \sum_{i=1,4} \sum_{j=1,2}
    G_{1}^{t\tilde{t}_j\tilde{\chi}_i^0} (D^{3,i,j}_{26}+D^{3,i,j}_{38}-
    D^{3,i,j}_{23} \\
&-& D^{3,i,j}_{39}) - \frac{i e^2 Q_b}{8 \pi^2} \sum_{i=1,2} \sum_{j=1,2}
    F_{1}^{t\tilde{b}_j\tilde{\chi}_i^+} (D^{4,i,j}_{25}+D^{4,i,j}_{35}-
    D^{4,i,j}_{23}-D^{4,i,j}_{37})
\end{eqnarray*}

\begin{eqnarray*}
f_{21,22}^{b,\hat{t}}=0.
\end{eqnarray*}
\par
The form factors in the renormalized amplitude of the quartic interaction
diagrams in Fig.1(d) have the form as:
\begin{eqnarray*}
f_1^{q} &=& f_2^{q} =
\frac{-i e^2}{32 \pi^2} \left[ Q_b^2\sum_{i=1,2} \sum_{j=1,2}(
    F_{2}^{t\tilde{b}_j\tilde{\chi}_i^+} m_{\tilde{\chi}_i^+} C^{1,i,j}_{0}-
    F_{1}^{t\tilde{b}_j\tilde{\chi}_i^+} m_t C^{1,i,j}_{11}) \right.\\
&+& Q_t^2\sum_{i=1,4} \sum_{j=1,2}(G_{2}^{t\tilde{t}_j\tilde{\chi}_i^0}
    m_{\tilde{\chi}_i^0} C^{2,i,j}_{0}-G_{1}^{t\tilde{t}_j\tilde{\chi}_i^0} m_t
     C^{2,i,j}_{11})\\
&+& 2 i Q_b^2\sum_{k=1,2}\bar{B}^{1,k}_{0} (A_h V_{h^0tt}
    V_{h^0\tilde{b}_k\tilde{b}_k}+A_H V_{H^0tt} V_{H^0\tilde{b}_k\tilde{b}_k})
    \\
&+& \left. 2 i Q_t^2\sum_{k=1,2}\bar{B}^{2,k}_{0} (A_h V_{h^0tt}
    V_{h^0\tilde{t}_k\tilde{t}_k}+A_H V_{H^0tt} V_{H^0\tilde{t}_k\tilde{t}_k})
     \right]
\end{eqnarray*}
$$
f_i^{q}=0,~(i=3 \sim 22)
$$
\par
The form factors in the renormalized amplitude from the t-channel triangle
diagrams depicted in Fig.1(e), are listed below:
\begin{eqnarray*}
f_1^{tr,\hat{t}} &=& f_2^{tr,\hat{t}} =
\frac{e^2}{8 \pi^2} \sum_{k=1,2} m_{\tilde{\chi}_k^+} \left[ 2 p_1 \cdot p_2
    C^{3,k}_{22}+(p_1 \cdot p_3+p_2 \cdot p_3) \cdot  \right.\\
&&  \left. (C^{3,k}_{0}-2 C^{3,k}_{23})+\epsilon \bar{C}^{3,k}_{24}+2 m_t^2
    C^{3,k}_{22}-m_{\tilde{\chi}_k^+}^2 C^{3,k}_{0} \right] \cdot \\
&&  (A_h V_{h^0tt} V_{h^0\tilde{\chi}_k^+\tilde{\chi}_k^+}^{s}+A_H V_{H^0tt}
    V_{H^0\tilde{\chi}_k^+\tilde{\chi}_k^+}^{s})-\left[
    \frac{e^2 Q_b^2}{8 \pi^2} \sum_{k=1,2} \bar{C}^{4,k}_{24} \cdot  \right.\\
&&  \left. (A_h V_{h^0tt} V_{h^0\tilde{b}_k\tilde{b}_k}+A_H V_{H^0tt}
    V_{H^0\tilde{b}_k\tilde{b}_k})+
    (Q_b, \tilde{b}, C^{4,k} \rightarrow Q_t, \tilde{t}, C^{5,k}) \right]
\end{eqnarray*}
\begin{eqnarray*}
f_7^{tr,\hat{t}} &=& f_8^{tr,\hat{t}} = f_9^{tr,\hat{t}} = f_{10}^{tr,\hat{t}} =
-\frac{e^2}{4 \pi^2} \sum_{k=1,2} m_{\tilde{\chi}_i^+} (C^{3,k}_{0}+4
    C^{3,k}_{22}-4 C^{3,k}_{23}) \cdot \\
&&  (A_h V_{h^0tt} V_{h^0\tilde{\chi}_k^+\tilde{\chi}_k^+}^{s}+A_H V_{H^0tt}
    V_{H^0\tilde{\chi}_k^+\tilde{\chi}_k^+}^{s})-\left[
    \frac{e^2 Q_b^2}{4 \pi^2} \sum_{k=1,2} (C^{4,k}_{23}-C^{4,k}_{22}) \right.
    \cdot \\
&&  (A_h V_{h^0tt} V_{h^0\tilde{b}_k\tilde{b}_k}+A_H V_{H^0tt}
    V_{H^0\tilde{b}_k\tilde{b}_k}) + \left. (Q_b, \tilde{b},
    C^{4,k} \rightarrow Q_t, \tilde{t}, C^{5,k}) \right]
\end{eqnarray*}
\begin{eqnarray*}
f_{21}^{tr,\hat{t}} &=& f_{22}^{tr,\hat{t}} =
\frac{-i e^2}{4 \pi^2} \sum_{k=1,2} m_{\tilde{\chi}_k^+} C^{3,k}_{0} (A_A
    V_{A^0tt} V_{A^0\tilde{\chi}_k^+\tilde{\chi}_k^+}^{ps}+A_G V_{G^0tt}
    V_{G^0\tilde{\chi}_k^+\tilde{\chi}_k^+}^{ps})
\end{eqnarray*}
$$
f_{i}^{tr,\hat{t}}=0,~~(i=3 \sim 6, 11 \sim 20),
$$
where $\bar{C}^{3,k}_{24}=C^{3,k}_{24}-\frac{\Delta}{4}$ and
$\bar{C}^{4,k}_{24}=C^{4,k}_{24}-\frac{\Delta}{4}$.
The form factors in renormalized amplitude of the self-energy corrections
${\cal M}^{s,\hat{t}}$ from Fig.1(f) in t-channel, are expressed as:
\begin{eqnarray*}
f_i^{s,\hat{t}} = 0,~~~~(i=2 \sim 4, 6 \sim 10, 12 \sim 22),
\end{eqnarray*}
\begin{eqnarray*}
f_1^{s,\hat{t}} &=&
\frac{-i e^2 Q_t^2 A_t^2}{16 \pi^2} p_1 \cdot p_3 \left\{
   16 \pi^2 (C^-_S + C^+_S - m_t C_L - m_t C_R) \right. \\
&+& \sum_{i=1,4} \sum_{j=1,2} \left[
       m_{\tilde{\chi}_i^0} G_{2}^{t\tilde{t}_j\tilde{\chi}_i^0}
       \bar{B}_0^{3,i,j} - m_t G_{1}^{t\tilde{t}_j\tilde{\chi}_i^0}
       \bar{B}_1^{3,i,j} \right] \\
&+& \sum_{i=1,2} \sum_{j=1,2} \left[ \left.
       m_{\tilde{\chi}_i^+} F_{2}^{t\tilde{b}_j\tilde{\chi}_i^+}
       \bar{B}_0^{4,i,j} - m_t F_{1}^{t\tilde{b}_j\tilde{\chi}_i^+}
       \bar{B}_1^{4,i,j} \right] \right\}
\end{eqnarray*}
\begin{eqnarray*}
f_{11}^{s,\hat{t}} &=& \frac{f_{5}^{s,\hat{t}}}{2} =
\frac{i e^2 Q_t^2 A_t^2}{16 \pi^2} \left\{
   16 \pi^2 \left[ m_t(C^-_S + C^+_S) + (p_1 \cdot p_3 - m_t^2)(C_L + C_R)
            \right] \right. \\
&+& \sum_{i=1,4} \sum_{j=1,2} \left[
       m_t m_{\tilde{\chi}_i^0} G_{2}^{t\tilde{t}_j\tilde{\chi}_i^0}
       \bar{B}_0^{3,i,j} + (p_1 \cdot p_3 - m_t^2)
       G_{1}^{t\tilde{t}_j\tilde{\chi}_i^0} \bar{B}_1^{3,i,j} \right] \\
&+& \sum_{i=1,2} \sum_{j=1,2} \left[ \left.
       m_t m_{\tilde{\chi}_i^+} F_{2}^{t\tilde{b}_j\tilde{\chi}_i^+}
       \bar{B}_0^{4,i,j} + (p_1 \cdot p_3 - m_t^2)
       F_{1}^{t\tilde{b}_j\tilde{\chi}_i^+} \bar{B}_1^{4,i,j} \right] \right\}
\end{eqnarray*}
\par
In this work we adopted the definitions of two-, three-, four-point one-loop
Passarino-Veltman integral functions as shown in reference\cite{s19} and
all the vector and tensor integrals can be deduced in the forms of scalar
integrals \cite{s20}.

\vskip 20mm

\vskip 20mm
\noindent{\Large\bf Figure captions}
\vskip 5mm

\noindent
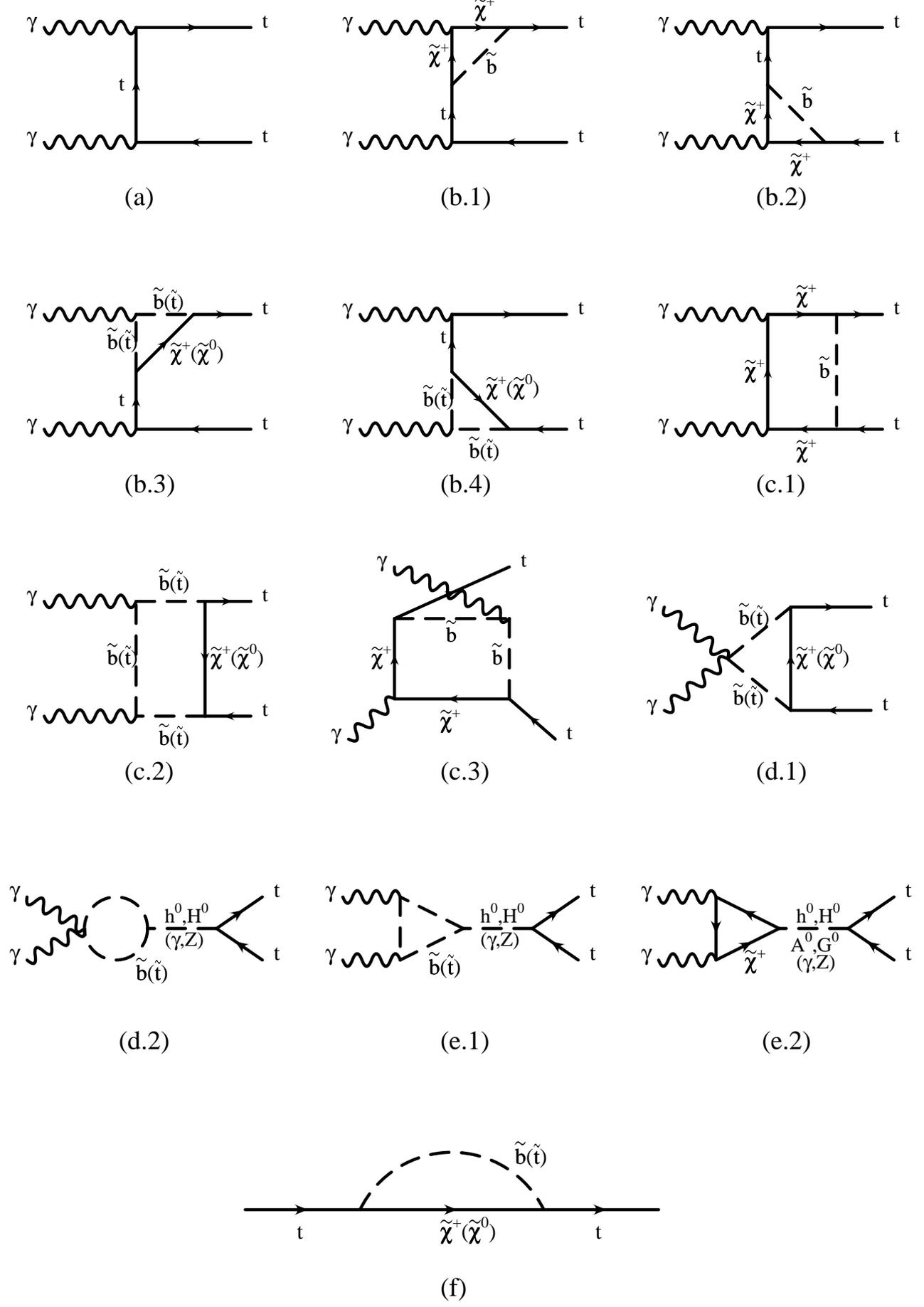
{\bf Fig.1} The Feynman diagrams at tree level and EW-like one-loop diagrams
        in the MSSM for subprocess $\gamma\gamma \rightarrow t\bar{t}$.
        (a) tree level diagram; (b) vertex diagrams; (c) box diagrams;
        (d) quartic coupling diagram; (e) triangle diagrams, and
        (f) self-energy diagrams. The $\tilde{t}$ and $\tilde{b}$
        that appear in diagrams have two physical particle eigenstates,
        while $\tilde{\chi}^{0}$ have four mass eigenstates, and
        $\tilde{\chi}^{+}$ have two. The diagrams with exchanging
        incoming photons are not shown in the figures except for Fig.1(d).

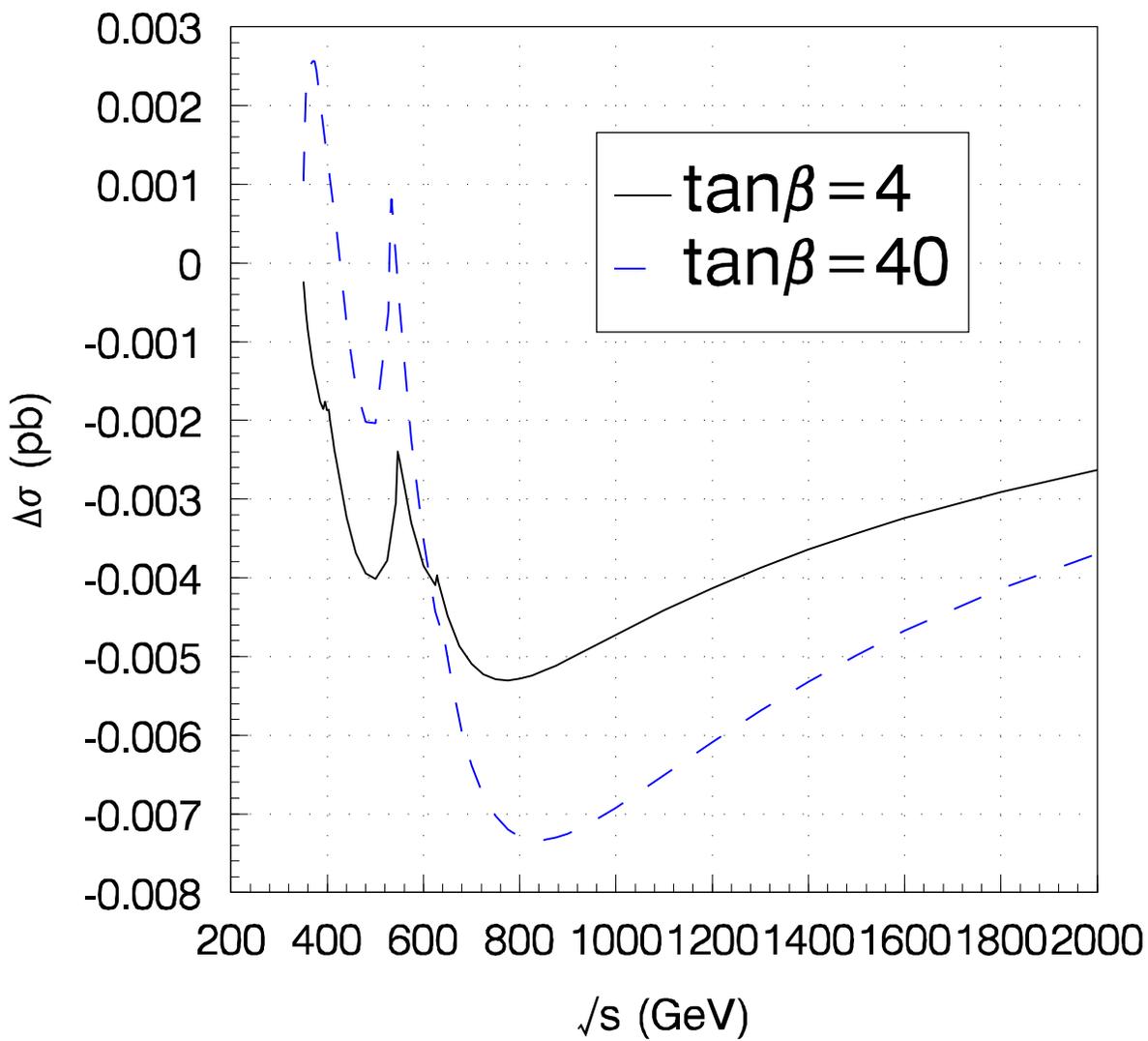
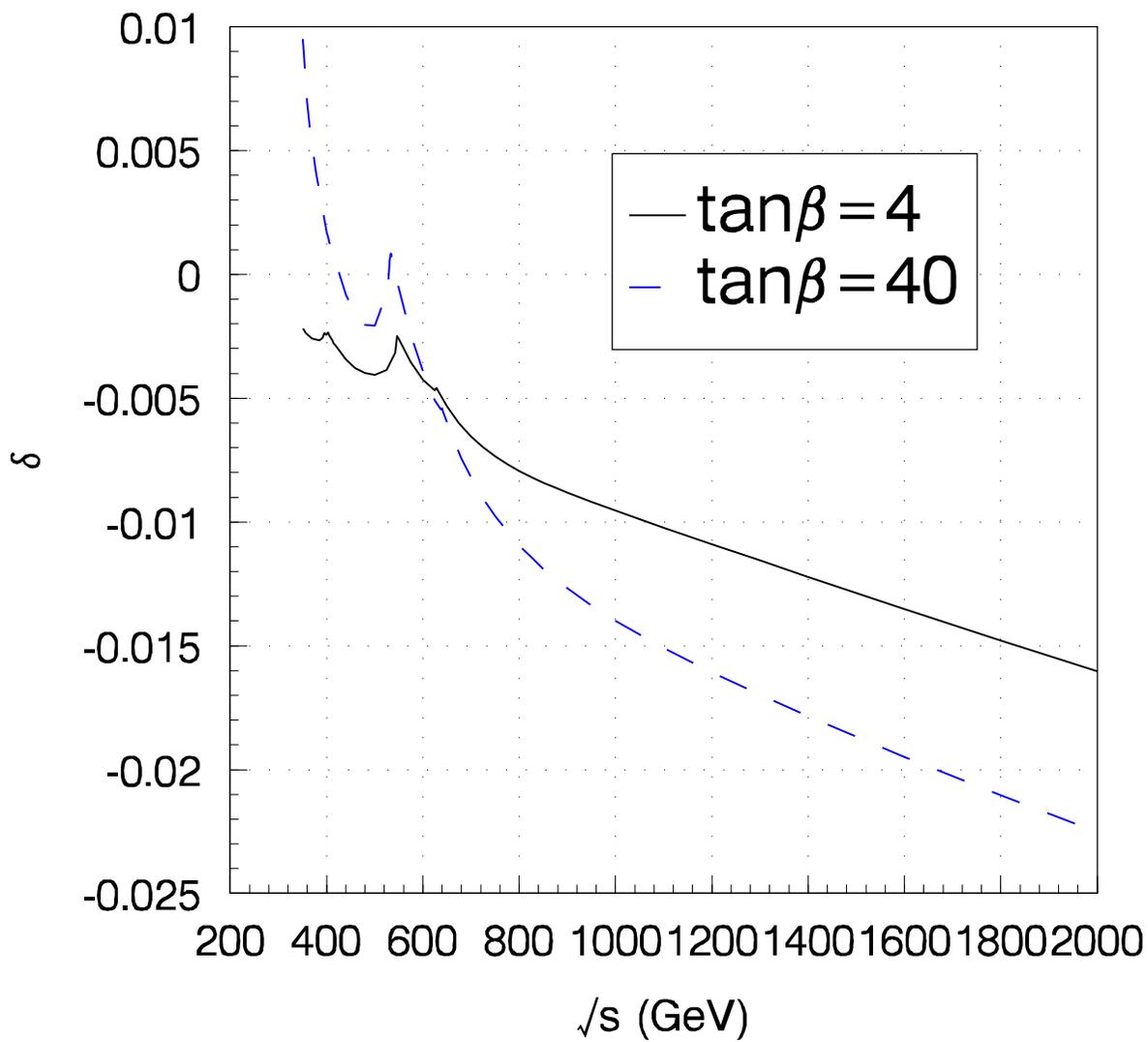
{\bf Fig.2} The corrections as the functions of c.m.s. energy $\sqrt{\hat{s}}$
        for subprocess $\gamma\gamma \rightarrow t\bar{t}$.
        (a) the absolute corrections; (b) the relative corrections.
        The solid line is for $\tan\beta=4$ and the dashed line is for
        $\tan\beta=40$.

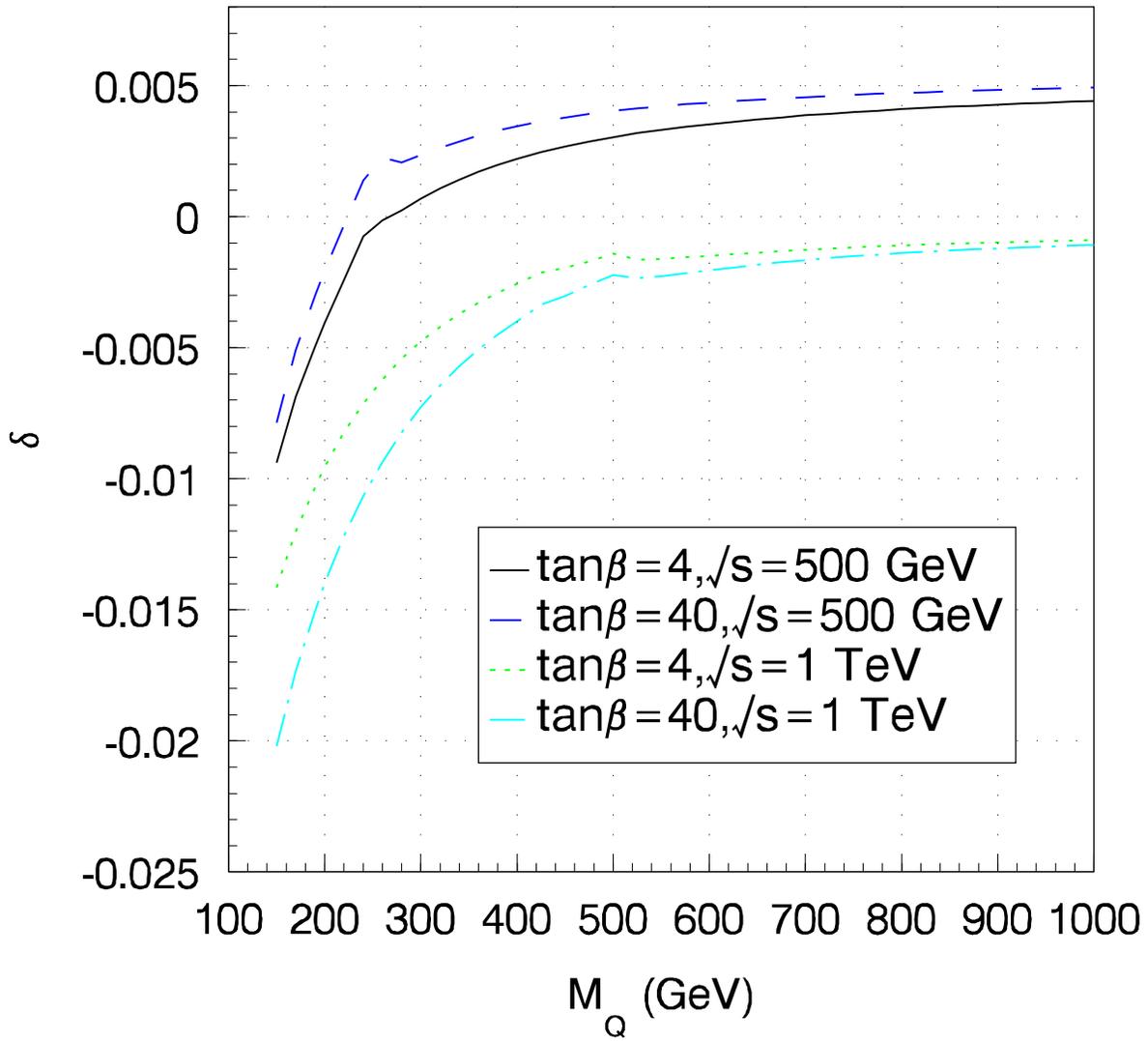
{\bf Fig.3} The relative corrections as the functions of $M_Q$
        for subprocess $\gamma\gamma \rightarrow t\bar{t}$.
        The solid line is for $\tan\beta=4,~\sqrt{\hat{s}}=500~GeV$,
        the dashed line is for $\tan\beta=40,~\sqrt{\hat{s}}=500~GeV$,
        the dotted line is for $\tan\beta=4,~\sqrt{\hat{s}}=1~TeV$, and
        the dash-dotted line is for $\tan\beta=40,~\sqrt{\hat{s}}=1~TeV$.

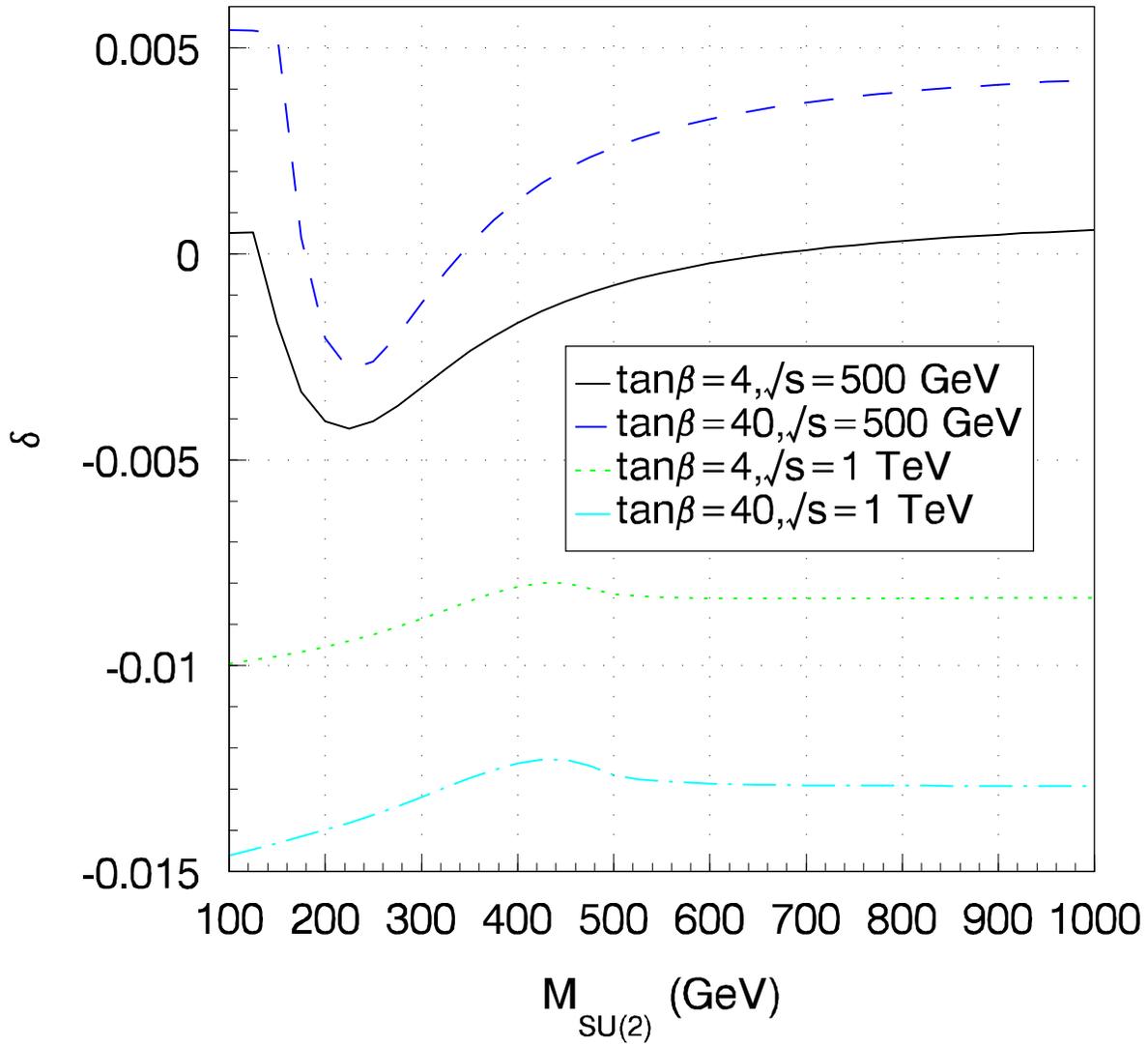
{\bf Fig.4} The relative corrections as the functions of $M_{SU(2)}$
        for subprocess $\gamma\gamma \rightarrow t\bar{t}$.
        The solid line is for $\tan\beta=4,~\sqrt{\hat{s}}=500~GeV$,
        the dashed line is for $\tan\beta=40,~\sqrt{\hat{s}}=500~GeV$,
        the dotted line is for $\tan\beta=4,~\sqrt{\hat{s}}=1~TeV$, and
        the dash-dotted line is for $\tan\beta=40,~\sqrt{\hat{s}}=1~TeV$.

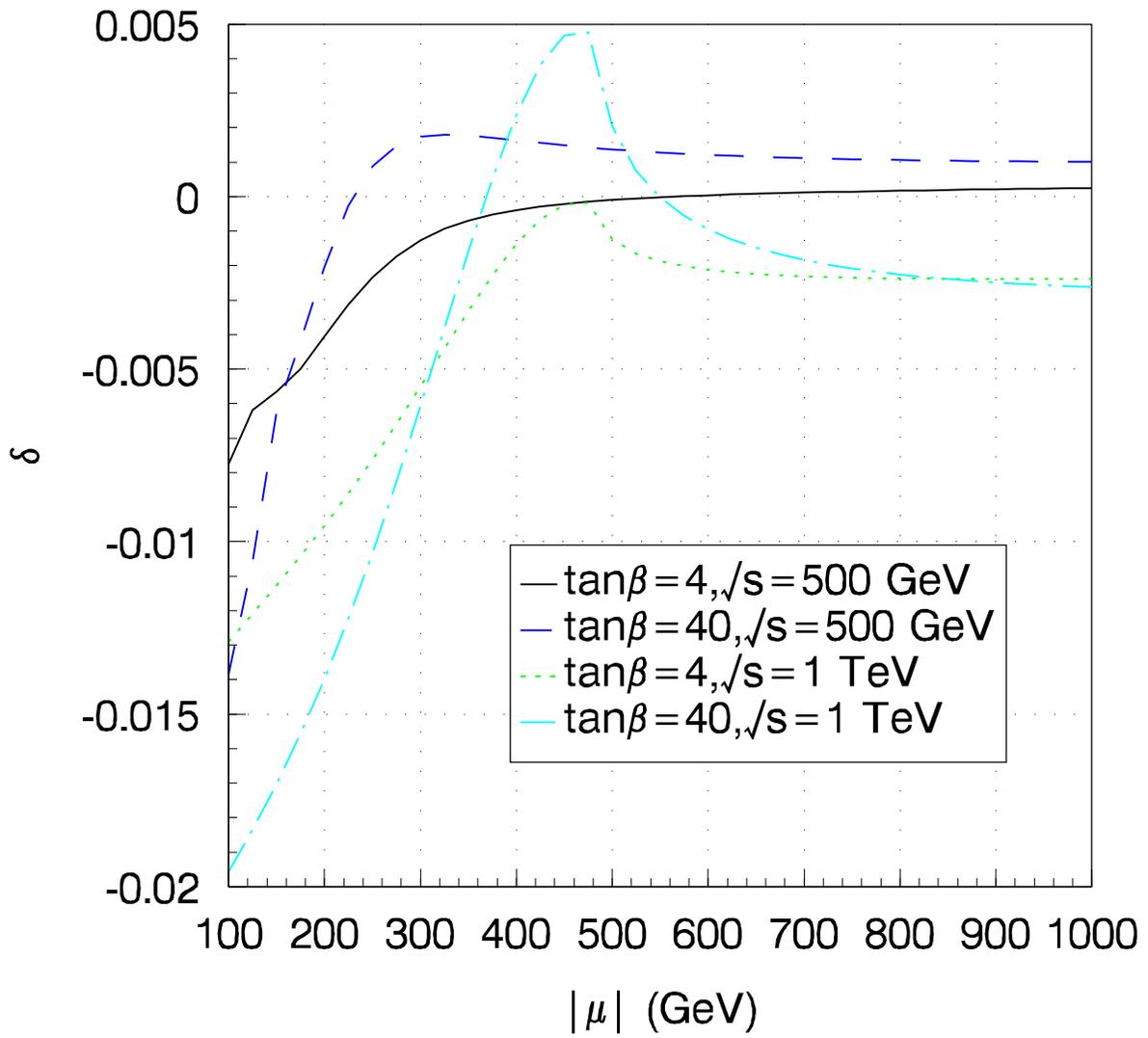
{\bf Fig.5} The relative corrections as the functions of $\mu$
        for subprocess $\gamma\gamma \rightarrow t\bar{t}$.
        The solid line is for $\tan\beta=4,~\sqrt{\hat{s}}=500~GeV$,
        the dashed line is for $\tan\beta=40,~\sqrt{\hat{s}}=500~GeV$,
        the dotted line is for $\tan\beta=4,~\sqrt{\hat{s}}=1~TeV$, and
        the dash-dotted line is for $\tan\beta=40,~\sqrt{\hat{s}}=1~TeV$.

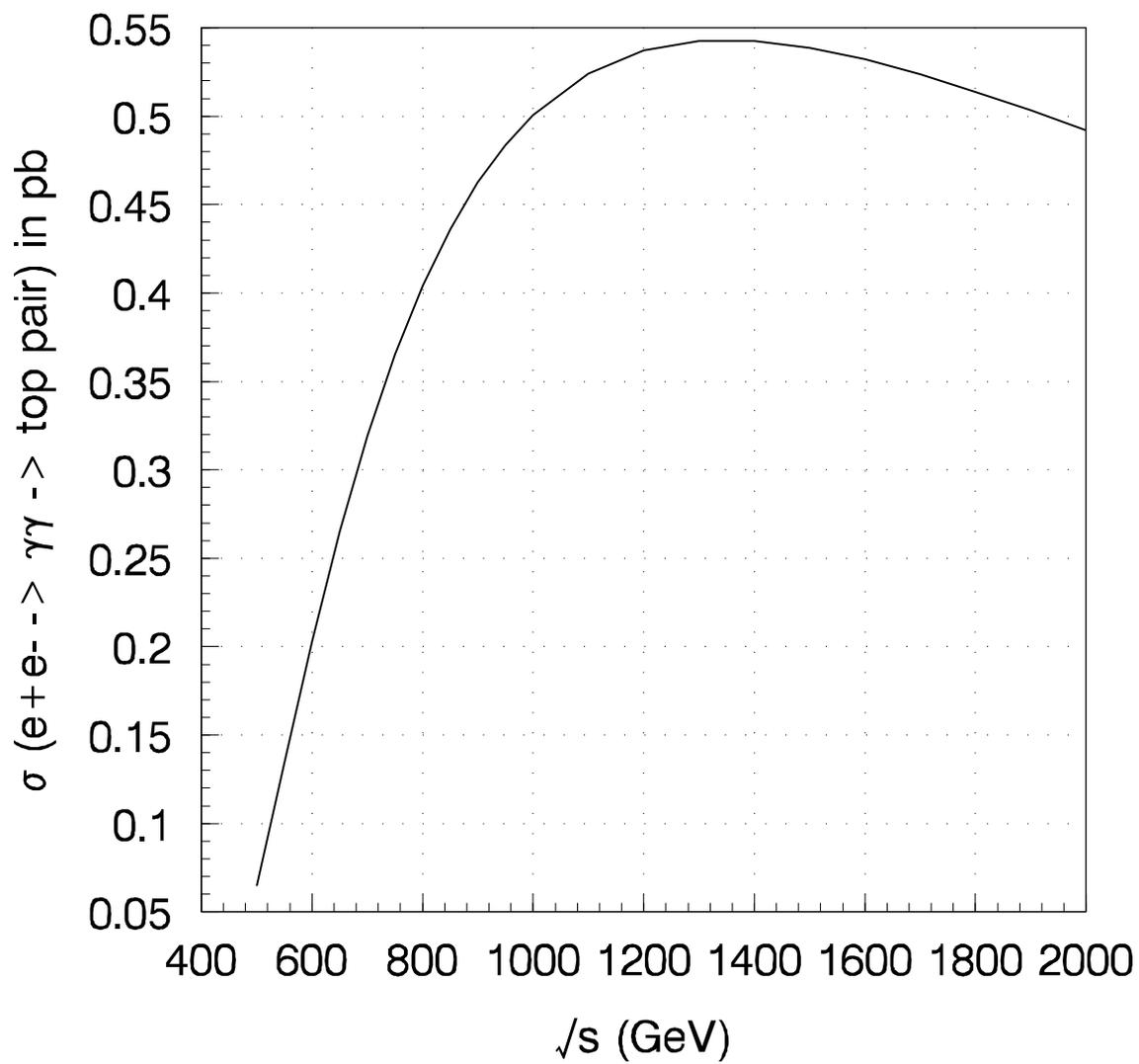
{\bf Fig.6} The cross section including the contributions of one-loop EW-like
        corrections for the parent process as the function of $e^+ e^-$ energy
        $\sqrt{s}$.

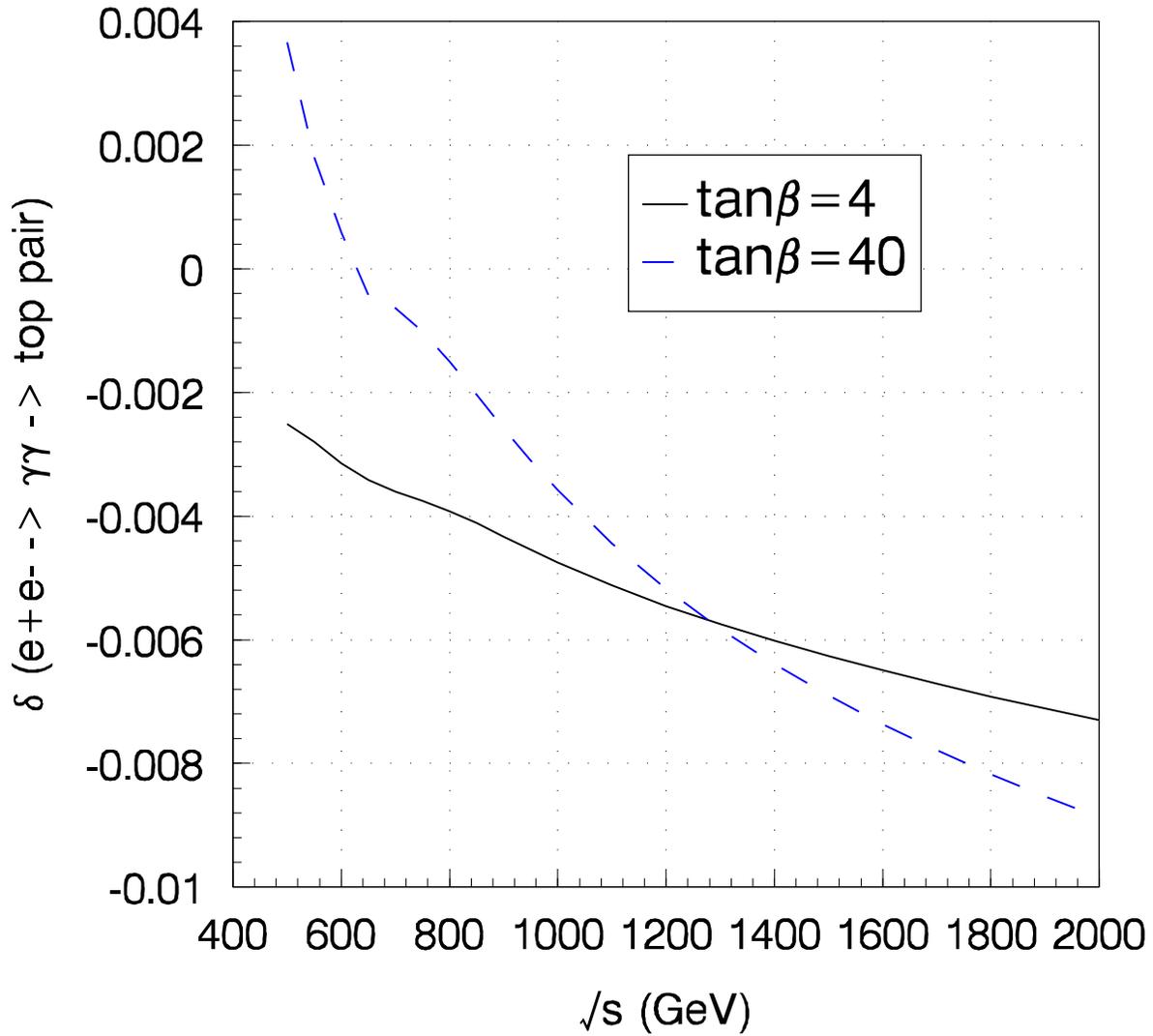
{\bf Fig.7} The relative correction for the parent process
        as the function of $e^+ e^-$ energy $\sqrt{s}$.
        The solid line is for $\tan\beta=4$ and the dashed line is for
        $\tan\beta=40$.

\vskip 3mm
\noindent

\vskip 3mm
\end{document}